\documentclass[fleqn,usenatbib]{mnras}

\usepackage{newtxtext} 
\usepackage[T1]{fontenc} 
\usepackage{ae,aecompl}

\usepackage{graphicx}
\usepackage{amsmath,amssymb}

\usepackage{newtxmath}

\usepackage[dvipsnames]{xcolor}
\title[Iron Rain: measuring the occurrence rate and origin of small iron meteoroids at Earth] {Iron Rain: measuring the occurrence rate and origin of small iron meteoroids at Earth}
\author[T. \ Mills et. al.]{
Tristan Mills$^{1,2}$\thanks{E-mail:tmills8@uwo.ca}, 
P. G.\ Brown$^{1,2}$, 
M. J. Mazur$^{1,2}$, 
D. Vida$^{1,2}$, 
Peter S. Gural$^{3}$
\newauthor
and Althea V.\ Moorhead$^{4}$  
\\
$^{1}$Department of Physics and Astronomy, University of Western Ontario, London, Ontario, N6A 3K7, Canada\\
$^{2}$ Institute for Earth and Space Exploration (IESX), The University of Western Ontario, London, Ontario N6A 3K7, Canada\\
$^3$Gural Software and Analysis LLC, Lovettsville, VA USA\\
$^{4}$NASA Meteoroid Environment Office, Marshall Space Flight Center EV44, Huntsville, Alabama 35812, USA%
}

\date{Accepted XXX. Received YYY; in original form ZZZ}

\pubyear{2021}

\begin{document}
\label{firstpage}
\pagerange{\pageref{firstpage}--\pageref{lastpage}}
\maketitle

\begin{abstract}
We report results of a four-year survey using Electron Multiplied Charged Coupled Device (EMCCD) cameras recording 34761 two-station video meteor events complete to a limiting magnitude of +6. The survey goal was to characterize probable iron meteoroids. Using only physical properties of the meteor trajectories including early peaking light curves, short luminous trajectories, and high energies accumulated per area at beginning, we identified 1068 iron meteors. Our iron candidates are most abundant at slow speeds $<15$ km/s, where they make up $\approx$20\% of the mm-sized meteoroid population. They are overwhelmingly on asteroidal orbits, and have particularly low orbital eccentricities and smaller semi-major axes when compared to non-irons between 10-20 km/s. Our iron population appears to be more numerous at fainter magnitudes, comprising 15\% of slow (10-15 km/s) meteors with peak brightness of +3 with the fraction rising to 25\% at +6 to +7, our survey limit. The iron orbits are most consistent with an asteroidal source and are in highly evolved orbits, suggesting long collisional lifetimes (10$^7$ years). Metal-rich chondrules (nodules) found in abundance in EL Chondrites are one possible source for this population. We also propose a possible technique using R-band colours to more robustly identify fainter iron meteors with very high confidence. 
\end{abstract}

\begin{keywords}
meteors -- meteoroids
\end{keywords}

\section{Introduction}
\label{sec:intro}

The composition and origin of extraterrestrial dust impacting Earth are varied. At large sizes, the many distinct classes of meteorites demonstrate directly that asteroidal parent bodies have experienced diverse evolutionary histories. The distinct populations of interplanetary dust particles and micrometeorites \citep{Bradley2003, Genge2008} also show that many different classes of solar system bodies deliver dust to Earth \citep{Carrillo-Sanchez2016} and that these parent populations differ significantly.

Among all the material impacting Earth, iron meteoroids/meteorites are among the most processed and distinct. While comprising 4\% of all observed meteorite falls \citep{Borovicka2019}, the fraction of apparently iron meteoroids producing fireballs is much smaller, of order 1\% \citep{Borovicka2006}. The difference is likely due to recovery biases, such as the higher density of iron meteorites, their resistance to weathering and their ability to be magnetically isolated. At still smaller sizes, the abundance of I-type micrometeorite spherules, believed to be derived from iron-dominated precursor meteoroids, recovered in Antarctica is $<2\%$ \citep{Taylor2000}. These results suggest that iron meteoroids are a small fraction of the impactor population at Earth. 

However, more recently, a spectral survey performed by \cite{Vojacek2015, Vojacek2019} classified 15 out of 152 meteors as iron at mm-sizes. \citet{Campbell-Brown2015} reported a population of low, slow meteors at mm and smaller meteoroid sizes that appeared to have properties consistent with iron, and appeared to increase in abundance with decreasing mass. This suggests that the fraction of iron meteoroids as a function of size and orbit type or entry speed may vary and raises the question: what fraction of the meteoroid population closer to the mass influx peak at Earth (near 10$^{-8}$ kg \cite{Love1993}) are iron meteoroids?

In addition to questions of the origin of such a unique compositional meteoroid population, `irons' pose a greater risk to spacecraft. Single-plate penetration equations can be used to assess how a spacecraft would withstand hypervelocity impacts, and many, such as the Cours-Palais equation, have a dependence on impactor density \citep{hayashida1991single}. Due to the higher density of iron, the predicted damage caused by iron meteoroids of similar velocity and mass is greater. Understanding this population of small iron meteoroids is thus also of great practical interest for both the design and operation of spacecraft.

Early detections of iron meteoroids were necessarily limited to bright and large meteors due to the insensitive nature of early photographic films. \citet{Halliday1960} reported a fireball spectrum coinciding with neutral iron that had a low velocity and begin height. The meteoroid was high density, and asteroidal in origin based on its orbit. \citet{ceplecha1966complete, ceplecha1967spectroscopic} reported a particularly large fireball with peak brightness of -10 magnitude, whose spectrum consisted of Fe I, Fe II, and trace Ni I and Co I lines. The authors theorized a spraying of droplets as the main process of ablation. Yet another study of brighter iron meteors identified 7 out of 287 fireballs as being caused by iron meteoroids at an initial mass range of 3-280 g \citep{ReVelle1994}, representing a roughly 2-5\% occurrence frequency in the tens of gram sizes across all speeds.

Evidence from direct collection at smaller sizes suggests irons may be more abundant. Iron micrometeorites (MMs) have been identified in the form of I-type spherules\textemdash oxidized iron grains \citep{Genge2008}. While these particles can provide insight into small iron particles near Earth, MMs, and particularly I-type spherules, suffer heavy biasing; I-types are resistant to weathering and can be separated from sediments via magnets, and are also more likely to survive atmospheric entry than silicate particles \citep{Genge2016}. Investigations of cosmic spherules report I-type abundances of 2\% from Antarctic wells, and close to 30\% from deep-sea collections \citep{Maurette1987,Taylor1991,Taylor2000}. Most I-type spherules are consistent with asteroidal sources, mostly from ordinary chondrite parent bodies \citep{Genge2016}.

The first comprehensive spectral survey of sporadic meteors with peak magnitudes between +3 and -1 magnitude (roughly cm-sizes), by \cite{Borovicka2005}, identified 14 of 97 meteors as having only iron emission. The authors suggested this population could reflect a pure iron, iron-nickel, or iron sulfide population. These iron-rich meteoroids displayed sudden onsets of luminosity with their peak brightness occurring in the first half of their trajectories. All but one had velocities less than 20 km/s, and all but one had begin heights lower than 90 km.

A larger video survey at similar sensitivities (mm-sizes) performed by \cite{Vojacek2015, Vojacek2019} found that 15 out of 152 meteors had characteristics consistent with iron meteoroids. In particular, these events possessed short, faint light curves, reached peak brightness early in their trajectories, became luminous at low heights, and were most common at low speeds. All of these characteristics are suggestive of a distinct population, a conclusion supported by the height and lightcurve differences as compared to spectrally ``normal'' meteors \citep{Borovicka2005, Vojacek2015}. \citet{Vojacek2015} noted that only representative shower spectra (across different brightnesses) were used in their analysis, and so the fraction of iron meteors still exceeds 2\% if the total number of spectra selected from was under 750 meteors.

Augmenting these data from mm-sizes, \citet{Matlovic2019a} presented a combined spectral-physical-orbital survey of 202 meteors having peak brightness from -1 to -14, corresponding to cm to dm sizes. Only one pure iron spectra was found in their sample, corresponding to an abundance of 0.5\%. This lone iron meteoroid was asteroidal, had high material strength but a short trajectory, and was cm-sized. The authors suggest that at cm to dm sizes irons are significantly less abundant than at mm sizes, which is supported by other studies that have probed to dimmer brightnesses and smaller sizes such as \citet{Campbell-Brown2015, Borovicka2005, Vojacek2015}. 

Another study at cm-sizes was performed by \citet{Vojacek2020}, who found 8 iron spectra out of their 220 meteor sample\textemdash an abundance of 3.6\%. These irons generally behaved similarly to the irons of \citet{Vojacek2019}.

In a larger study of 7000 two-station video events with limiting meteor magnitudes of +6.5, a distinct ``low, slow population'' of meteors with velocities below 30 km/s and begin heights less than 86 km was found \citep{Campbell-Brown2015}, though no spectral information was available. These meteors also displayed short light curves, sudden onsets with early peaks, and began at low heights for their speeds, consistent with the iron characteristics identified by \cite{Borovicka2005, Vojacek2015}. These low, slow, sudden onset meteors also became more prominent at fainter magnitudes; 3\% of meteors brighter than +3 showed these characteristics while between peak meteor magnitudes of +6 and +3 more than 6\%  had these features. 

Motivated in part by these contemporary observations, \citet{Capek2017} developed a comprehensive ablation model for small iron meteoroids. The authors tested several ablation scenarios for irons guided by the existing data on begin heights, trail lengths, and light curve shapes validated with meteors previously identified through spectra as being iron. They found best agreement using a model whereby ablation is dominated by melting and droplet injection into the air stream. 
 
Building on this early model work, \citet{Capek2019} improved the iron ablation model by considering how droplets ablate independently after a delay is introduced from their release into the air stream. This study examined 1500 two-station events of mm-sizes and isolated a population with short light curves and sudden luminosity onset. This low, slow population consisted of 45 possible iron meteors, 9 of which were confirmed irons based on spectra. The improved iron ablation model reasonably described these meteors in terms of begin height, brightness, and length, though became less consistent with measurements for larger (brighter) iron meteors. A comprehensive hydrodynamic model of meteoroid melting and ablation has also been presented by \citet{Girin2019} and produces similar observational predictions to \citet{Capek2019}. 

\cite{Vojacek2019} noted that iron meteoroids typically require more energy per unit cross-section prior to become visible than stony material. This is because not only is heat better conducted throughout the body of an iron meteoroid, but iron's heat of vaporization is also very high. As a result, irons must receive higher total amounts of energy to start spraying droplets, and thus will also penetrate deeper into the atmosphere \citep{Genge2016}. 

Most of the meteoroids in the \citet{Capek2019} sample possess asteroidal orbits, with inclinations lower than 40$^{\circ}$. All 9 of the detected two-station irons from \cite{Borovicka2005}, 10 of 15 from \cite{Vojacek2019}, and 7 of 8 from \cite{Vojacek2020}, had asteroidal orbits with inclinations less than 45$^{\circ}$, aphelia in the asteroid belt at less than 4.5 AU, and Tisserand parameters with respect to Jupiter greater than 3. One iron each from \cite{Vojacek2019} and \cite{Vojacek2020} was found to have a perihelion smaller than 0.2 AU and was classified as sun-approaching. Of the remaining irons from \cite{Vojacek2019}, two were Jupiter-family objects, with inclination smaller than 45$^{\circ}$, aphelia larger than 4.5 AU, and Tisserand parameters with respect to Jupiter between 2 and 3. They note, however, that an asteroidal origin for these Jupiter-family irons cannot be excluded. Interestingly, there were also two irons with Halley-type orbits with inclination larger than 60$^{\circ}$. It is unclear how they came to these orbits other than having originated from HTCs. 

The low, slow population identified by \citet{Campbell-Brown2015} also generally possessed asteroidal orbits. While the Tisserand parameters and low inclinations of these objects suggests asteroidal origins, the possibility that they evolved through radiative forces into these orbits from more cometary-like orbits cannot be ruled out. 

One scenario is that the population originally possessed Jupiter-family type orbits, which then shrank and circularized through Poynting-Robertson drag, and dynamically decoupled from Jupiter. The timescale to move from $T_\mathrm{J}<3$ to $T_\mathrm{J}>3$ would be of order 10$^{5}$ years under PR effects \citep{Subasinghe2016} for mm-sized iron meteoroids. The moderate eccentricities of the small number of published low, slow meteoroids with iron ablation characteristics suggest, however, that PR drag has not had a dominant effect on these orbits, though they tend to be more circular than the remaining population. The variety of orbits seen in irons of which there are spectra, be they Jupiter-family, sun-approaching, or Halley-type irons, point to complex orbital evolutions \citep{Vojacek2019}.

The goal of the present study is to constrain the fraction of iron meteoroids in the mm size range impacting the Earth, and to explore their likely origin. In contrast to prior studies, we focus on the smallest meteoroid population detectable with optical instruments using a statistical\textemdash rather than event-by-event\textemdash approach. In particular, we wish to survey the abundance of irons as a function of entry speed, size/brightness and orbit-type, thereby requiring significant number statistics. 

To accomplish this goal, we analyze 34761 two-station video meteors detected down to a limiting peak magnitude of +7 (with measured light curves as dim as +8) that were recorded during a four-year automated survey. Motivated by the work of \cite{Capek2019}, we develop selection criteria based on the ablation behaviour of individual meteors to isolate probable iron meteoroids. These criteria include: comparatively low begin heights as a function of velocity, short trajectories, early brightness peaks, and limiting values for intercepted air stream kinetic energy per unit cross-sectional area.

Section \ref{sec:equipment} summarizes the observational methods used to detect meteors, and briefly outlines how data processing is performed on video data. In Section \ref{sec:distinguishing} we further outline the ways in which we distinguish iron meteoroids from the general population. In Sections \ref{sec:results} and \ref{sec:discussion} we discuss the various properties of our selected iron candidates, including abundances and orbital characteristics, and address the question of detection biases.

\begin{table*}
	\centering
	\caption{Summary of literature estimates of Iron meteoroid abundances with size/magnitude/mass.}
	\label{tab:iron_abundances}
	\begin{tabular}{llll}
		\hline
		 Reference & Size/mag/mass & Iron Fraction & Technique\\
		\hline
		 \citet{Maurette1987} & $>50$ $\micron$ & 2\% & Greenland ice cap melt zone\\
		 \citet{Taylor1991} & $>100$ $\micron$ & 1\% & Greenland ice cap\\
		 \citet{Taylor1991} & $>100$ $\micron$ & 26\% & Deep sea collection\\
		 \citet{Taylor2000} & $>50$ $\micron$& 2\% & Antarctic well collection \\
		 & $>100$ $\micron$ & 1\%\\
		 \citet{ReVelle1994} & $10<\mathrm{m}<280$ g & 3\% & Photographic networks\\
		 \citet{Borovicka2005} & $1<\mathrm{s}<10$ mm & 14\% & Spectral survey\\
		 \citet{Campbell-Brown2015} & $\mathrm{M}<$ +3 & 3\% & Video survey \\
		 & +3 $<\mathrm{M}<$ +6 & 6\%\\
		 \citet{Vojacek2015, Vojacek2019} & -5 $<\mathrm{M}<$ +3 & 9.9\% & Spectral and video survey\\
		 \citet{Matlovic2019a} & -14 $<\mathrm{M}<$ -1 & 0.5\% & Spectral and video survey\\
		 \citet{Vojacek2020} & 1 $<\mathrm{s}<$ 4 cm & 3.6\% & Spectral and video survey\\
			\hline
	\end{tabular}
\end{table*}

\section{Equipment and Data Collection}
\label{sec:equipment}

\subsection{Optical Cameras}

\begin{figure}
    \centering
    \includegraphics[width=\linewidth]{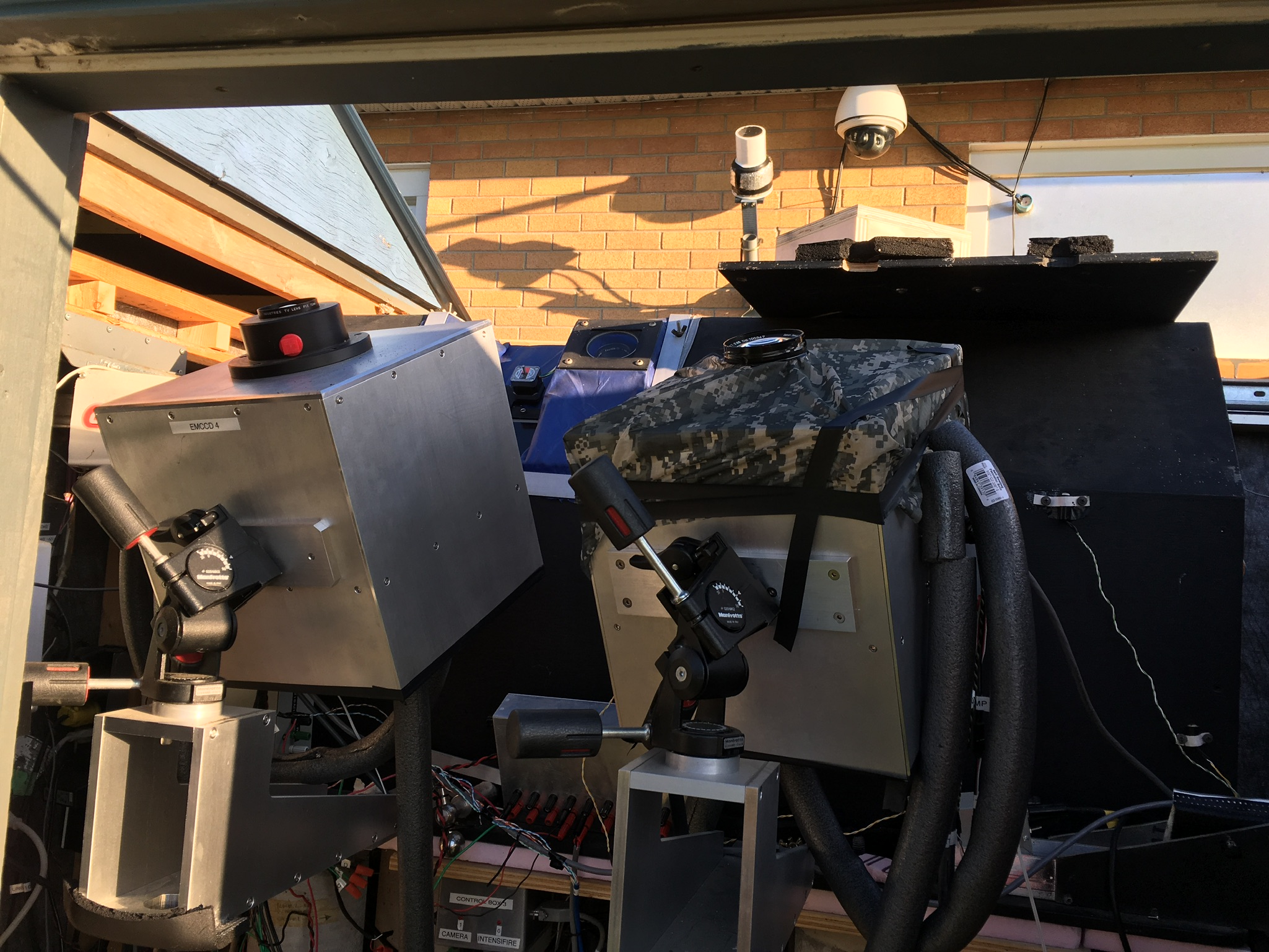}
    \vspace{-0.5cm}
    \caption{Two EMCCDs located at Elginfield Observatory (encased in aluminium housings for temperature and dust control).}
    \label{fig:emccds_insitu}
\end{figure}

To extend our survey to the faintest possible optical meteor magnitudes, where earlier surveys hint at an increase in the fraction of iron-rich meteoroids \citep[e.g.][]{Kaiser2005a}, we use visible-band, low-light sensitive Electron Multiplied Charged Coupled Devices (EMCCDs), pictured in Figure \ref{fig:emccds_insitu}. As we require trajectories, orbits, and complete lightcurves, we use a two-station, automated setup with four EMCCDs total, which are paired to overlap at moderate (100 km) and lower (80 km) heights. 

EMCCDs function similarly to traditional CCD detectors, but upon detection of photoelectrons, secondary electrons are produced. The signal is multiplied and read noise no longer contributes to the signal \citep{tulloch2011}. Moreover, EMCCDs have low dark current and can operate at higher clock frequencies than CCDs. The EMCCD cameras use a back-illuminated sensor, leading to very high quantum efficiency ($\approx$95\%). The use of EMCCDs in astronomical applications allows for photon counting of dim sources at high speeds\textemdash an optimal combination for fast meteor photometry.

\begin{table}
	\centering
	\caption{EMCCD Camera stations. See text for more details.}
	\label{tab:emccd_specs}
	\begin{tabular}{lc}
		\hline
		  & Specification\\
		\hline
		Location 1 (Lat., Long.) & 43.264$^{\circ}$N, 80.772$^{\circ}$W\\
		Location 2 (Lat., Long.) & 43.194$^{\circ}$N, 81.316$^{\circ}$W\\
		Pointing 1F/2F (Alt., Az.) & 64.9$^{\circ}$, 322.8$^{\circ}$ / 65.8$^{\circ}$, 6.6$^{\circ}$\\
		Pointing 1G/2G (Alt., Az.) & 47.0$^{\circ}$, 329.7$^{\circ}$ / 47.2$^{\circ}$, 359.7$^{\circ}$\\
		Camera & N\"{u}v\"{u} HN\"{u}1024\\
		Sensor & Teledyne e2v CCD201-20\\
		Pixels & 1024x1024 13$\mu$ m\\
		Digitization & 14-bit\\
		Readout Noise & $<0.1 e^{-}$\\
		Dark Current @ -85$^{\circ}$C & 0.0004 e$^{-}$/pixel/s\\
		EM Gain & up to 1000 (200 operational)\\
		Quantum Efficiency & ca. 95\%\\
		Framerate (1x1/2x2) & 16.7/32.7 fps\\
		Lens & Nikkor 50 mm f/1.2\\
		Field-of-View & 14.7$^{\circ}$x14.7$^{\circ}$\\
		Plate Scale (1x1/2x2) & 51.7"/103.4" per pixel\\
		Resolution at 100 km (1x1/2x2) & 25/50 m\\
		Stellar Magnitude Limit & +10.5\\
		Meteor Peak Magnitude Limit & +7.0\\
		\hline
	\end{tabular}
\end{table}

Our camera system consists of four N\"{u}v\"{u} HN\"{u} 1024 EMCCD cameras situated across two sites: Elginfield (43.264$^{\circ}$N, 80.772$^{\circ}$W) and Tavistock (43.194$^{\circ}$N, 81.316$^{\circ}$W) in southwestern Ontario, Canada. The EMCCDs were operated at 16.7 fps in 1x1 binning (1024x1024 pixels) between January 2017 and March 2018, then 32.7 fps at 2x2 binning (512x512 pixels) after March 2018. All imagery is 14-bit depth. Camera specifications are given in Table \ref{tab:emccd_specs}. The cameras begin to show saturation effects near meteor magnitudes of +1. For example meteor images and more details about the system see \citet{vida2020new}.

When the transmission of lenses used (Nikkor 50 mm f1.2) is convolved with the camera's response function, the system has a usable wavelength range of 350 nm to 900 nm, with peak sensitivity around 550 nm. The stellar magnitude limit per frame is +10.5, and the meteor magnitude limit is between +8 to +9, depending on angular velocity.

\begin{figure}
    \centering
    \includegraphics[width=\linewidth]{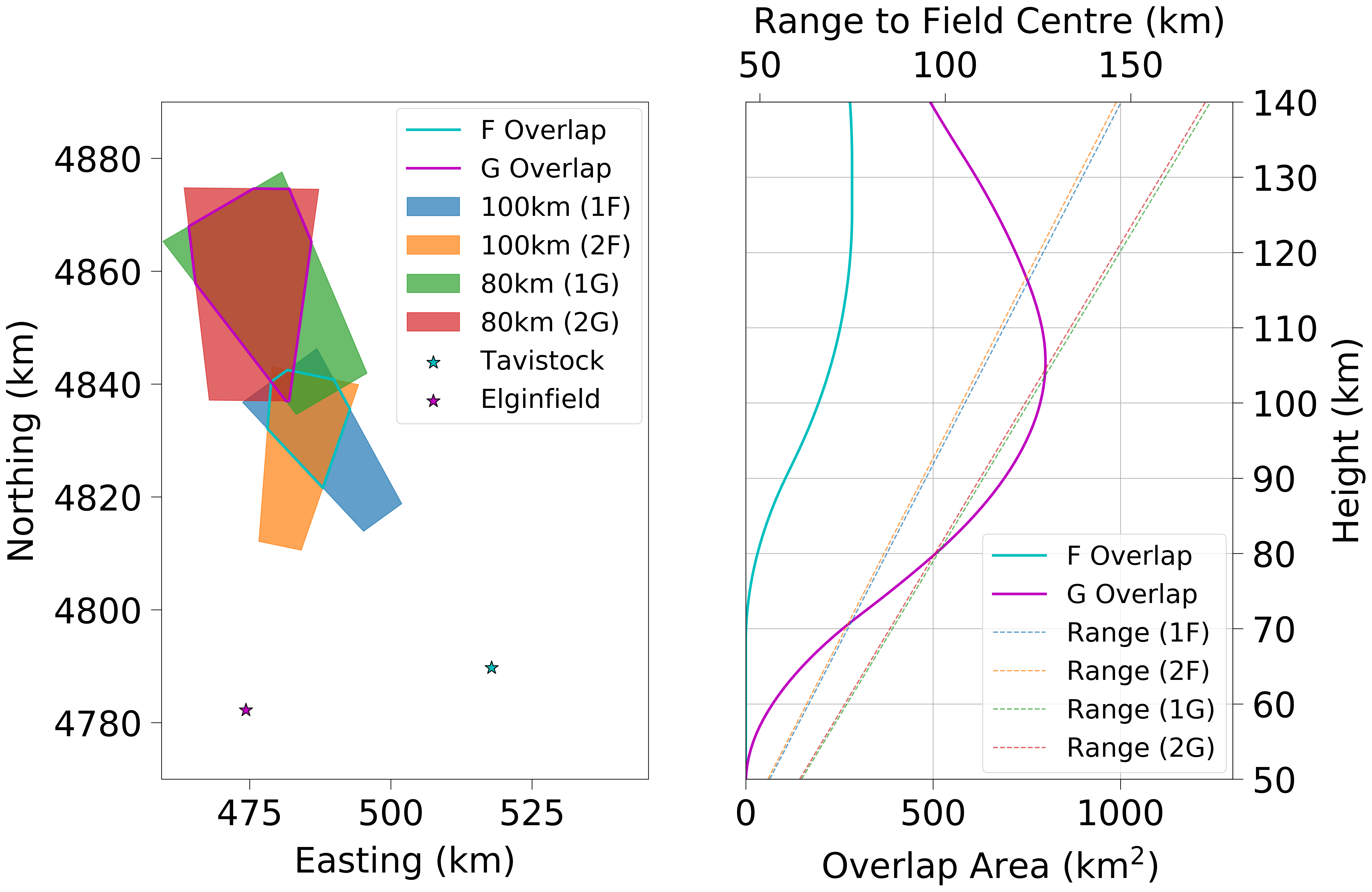}
    \vspace{-0.5cm}
    \caption{Field coverage projected onto a UTM grid in the atmosphere at 100 km height for the F camera pair, and 80 km for the G camera pair from Elginfield (leftmost station represented by star) and Tavistock (rightmost location) (left sub-plot). In the right sub-plot, the overlap areas as a function of height is shown for EMCCD camera pairs as well as the apparent range to the field centre at each height.}
    \label{fig:height-overlap}
\end{figure}

The EMCCDs are paired such that two fields of view are both observed from across the two sites as shown in Figure \ref{fig:height-overlap}. One pair of cameras is high pointing (labelled F) and the other is low pointing (labelled G). The F cameras' overlapping field of view has 150 km$^2$ of overlapping area at an altitude of 100 km while the G cameras cover over 200 km$^2$ of overlapping area between 70 km to $>120$ km. All EMCCD cameras are pointed north to avoid illumination from the moon and light pollution from population centres. 

As one instrument suite in the Canadian Automated Meteor Observatory (CAMO) system, the EMCCD cameras are autonomous. As part of normal operations, a separate guide camera monitors the  instrumental magnitude of Polaris with a cadence of 10s to detect cloudy conditions. When conditions are optimal (no moon, cloud, rain, high wind, or twilight), observations with the EMCCDs commence and are stored and automatically processed. More information on the CAMO system automation control and weather monitoring can be found in \citet{Weryk2013, Vida2021}.

We apply both flats and biases to our imagery to obtain light curves. We have observed that flat-field images do not change significantly with time (on scales of months) and so they are updated manually as needed. Astrometric pointing may drift slightly over much shorter periods of time than the flat and bias images. Automatic astrometry plates are generated and used for processing with a cadence of 10 min each night for each camera.

\subsection{Data Processing}

All raw image data gathered by the cameras each night is stored in a large ring buffer. Each camera records 10 minute blocks of video throughout the night and these base files are processed once data collection finishes in the morning. Data rates are 1 GB/min per camera. Detection of meteors is performed using a hybrid approach consisting of a fast front-end clustering/tracking algorithm that cues a matched filter measurement refinement and false alarm mitigation process, collectively called DetectionApplication (DetApp) (Gural, in preparation). 

Per frame output from DetApp includes a positional measurement estimate of the leading edge of meteor trails, the background subtracted summed intensity of the visible trail, and a maximum likelihood estimate from the matched filter processing. Only events with at least 4 detectable frames are retained. Apparent frame to frame motion must be at least 2 pixels (to exclude satellites) but no more than 150 pixels (maximum apparent angle rate for meteors given the focal plane characteristics). 

For our data pipeline, the individual camera photometric offsets are automatically computed at the time of each event and applied to each meteor's lightcurve. A candidate detection is declared if a user defined excursion of clustered pixel points above the background is reached (typically 1.5$\sigma$ above the mean). A matched filter refinement is then applied to define the orientation and length of the trail along the meteor's multi-frame track. The matched filter template requires the use of a motion propagation model. We have chosen a constant acceleration motion model across the focal plane to handle the geometric effects of foreshortening or lengthening from viewing geometry changes, and some level of meteor deceleration if present. No empirical model can mimic true meteor deceleration, but for these narrow field of view systems, the bulk of the velocity change is geometric. The resultant fit is made to the actual meteor trail per frame including an estimated and convolved point spread function, giving a best estimate of the leading edge of the meteor. Finally, the meteor track is extended at the start and end of the track such that all pick points above background noise are included in the photometric light curve.

Comparison of the resulting speeds with manual reductions shows good agreement, generally better than 1 km/s at typical meteor speeds (30 km/s). This is far below the level at which orbit types are likely to switch between major parent populations (HTC, JFC, asteroidal), our main concern in this study. However, care should be taken if these automated speeds are used for high precision orbit measurements - in those cases comparison with manual reductions should be undertaken. Data were collected in this way through the end of 2020 with the survey beginning in Jan 2017.

\begin{figure}
    \centering
    \includegraphics[width=\linewidth]{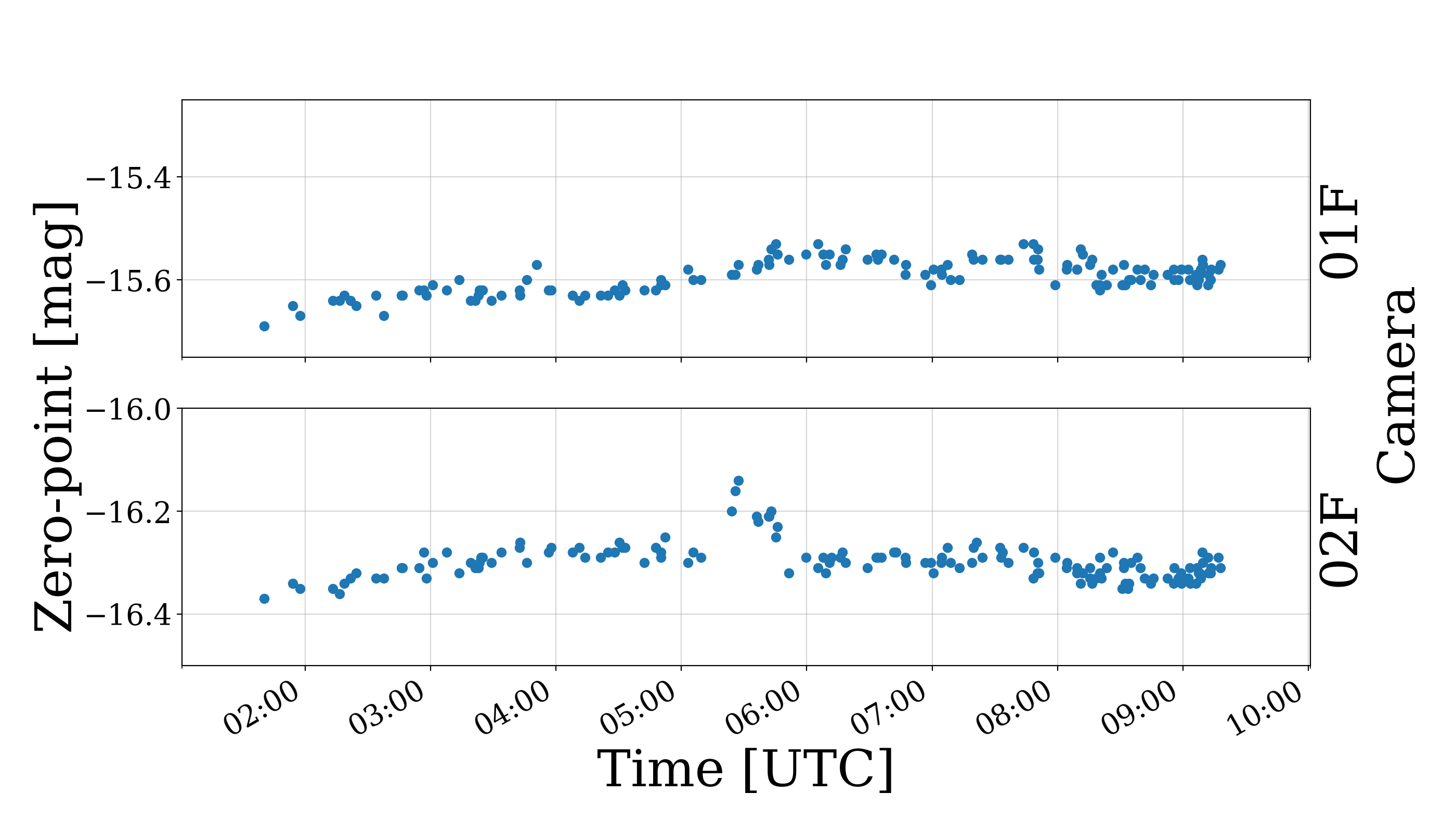}
    \vspace{-0.5cm}
    \caption{An example of the time change of photometric magnitude zero points of two paired cameras at different sites over the course of a single night (August 21st, 2020). Here each symbol represents a detected meteor, whose zero-point is dynamically computed per meteor event. All reference magnitudes for calibration are from the Gaia catalogue \citep{2016GaiaMission}.}
    \label{fig:phot_offsets}
\end{figure}

Due to the large volume of data captured and processed for each night of EMCCD video, only detected event video snippets are saved to long term storage. Median images of the begin and end time around meteor events were used to find the photometric offsets from this save-limited data. In total sixty frames of data bracketing the begin and end of an event (roughly 2 sec) are meteor-free and close enough in time that stars can be measured and compared against the Gaia G-band magnitude to obtain an accurate photometric offset per event. In the situation where an offset was unable to be computed for a particular event the next closest offset in time on that camera and date was used. Figure \ref{fig:phot_offsets} shows an example of the zero-point photometric drift computed this way on a night with good seeing.

Astrometric measurements are done using an affine automated plate fit to third order between focal plane x,y coordinates and Gaia G catalogue star positions, updated every ten minutes. Typical rms residuals are less than 0.01$^{\circ}$. Details of the plate transform and fit approach can be found in \citet{WerykBrown2012}.

Individual camera events at each site were correlated across sites to find true meteors. Initial cross-site correlations were grouped by time (with an acceptance window of 2 sec). A secondary filter removed any meteors which were found to have begin heights above 160 km or below 60 km, end heights above 120 km or below 50 km, and initial velocities above 90 km/s or below 6 km/s. The orbital and trajectory solutions for the remaining "good" meteors were computed using WesternMeteorPyLib (Pylig) \citep{vida2020mctheory} \footnote{Code accessible from: https://github.com/wmpg/WesternMeteorPyLib}. In total, 55707 two-station video meteors were computed after applying our initial filters. This number is lowered to 34761 when only considering meteors whose beginnings and endings are observed in at least one camera, which ensures full light curves and accurate trajectory lengths as the entire trajectory has been observed to our limiting sensitivity (see Section \ref{sec:distinguishing}).

However, to ensure data quality with respect to orbit measurement, an additional check was incorporated. During detection, if a meteor has too few points to fit a constant acceleration model with the matched filter, it is assumed that the meteor has a constant apparent angular velocity over such a short length. These short events can result in positive spatial accelerations in their Pylig solutions due to perspective effects. For these specific events, the number of points required to fit a constant acceleration was lowered and the trajectories and orbits were recalculated. Only events that initially reported positive spatial accelerations were reprocessed this way; other erroneous trajectories can occur when a short approximately constant velocity meteor is assumed to be decelerating. About 7400 meteors were flagged as accelerating and reprocessed again. In most of these cases only one station initially reported acceleration. The resulting change to orbital elements in most cases was quite small, but the correction applied nonetheless.

An additional complication in computing trajectories is algorithm specific. Pylig compares the velocities of multiple sites to make sure that they are consistent at each point in time. It therefore requires that a minimum of four overlapping points of a meteor's trajectory are observed between the stations. If there is a gap between the portions of a meteor that Pylig observes, Pylig is unable to find a solution. As a check on the Pylig solutions, a constant velocity model \citep{Gural2012} was also employed. 

The Gural solution uses an empirical meteoroid velocity model, so it is able to solve trajectories with interruptions in the observation. However, such a velocity model will not perfectly match observed behaviour and makes assumptions about the dynamic behaviour of the meteoroid \citep{egal2017challenge}. For this reason, after reprocessing as many events using Pylig as possible, the Gural solutions were used in cases where Pylig failed to match any timing offsets and such events were flagged. In general, comparison of the Pylig and constant velocity model fits for common meteor events produced very similar initial speeds, so this trajectory substitution for cases where two sites did not have overlapping time of observation is expected to produce only minor changes to the initial orbital elements. 

\section{Distinguishing Iron Meteors}
\label{sec:distinguishing}

The best means to distinguish iron meteoroids from among the general meteoroid population is through the use of spectra \citep{Borovicka2005}. This produces a robust means of distinguishing pure (or primarily) iron material from chondritic-like objects. However, for a given sensitivity instrument, spectral measurements are only possible on comparatively bright meteors. 

Past spectral surveys, for example, \citep[e.g.][]{Borovicka2005, Vojacek2015} are typically limited to meteors of peak brightness +2 or brighter, with the majority of meteor spectra brighter than this limit and hence appropriate to meteoroids significantly larger than mm in size. To probe the fraction of irons at the faintest possible optical magnitudes precludes spectral methods. Instead, we adopt quantitative metrics based on the shape of the lightcurve, beginning height of ablation, and trail length to isolate probable iron meteoroids. These criteria are adopted based on prior work \citep{Capek2019} where simultaneous spectra were available to truly distinguish irons and therefore empirically and theoretically calibrate these quantities. The iron ablation model proposed in that work best described smaller irons with magnitudes fainter than zero, appropriate to our study.


One metric of a meteor's ablation behaviour is the F parameter, which is the ratio of the distance between the begin height and height of peak brightness to the distance between the begin and end heights (total height extent of the trail). This can be expressed as:
\begin{align}
    F_\mathrm{H} &= \frac{H_\mathrm{B} - H_\mathrm{peak}}{H_\mathrm{B} - H_\mathrm{E}}
\end{align}
where $H_\mathrm{B}$, $H_\mathrm{E}$, $H_\mathrm{peak}$ are the begin height, end height, and height of peak brightness, respectively \citep{Koten2006}.

F parameter values range from 0 to 1, with 0.5 representing symmetrical light curves which peak in the middle. Ideal, single-body meteors tend to have late-peaking lightcurves with F parameters around 0.7, and dustball-type meteors are more symmetric with F parameters of 0.5 \citep{ceplecha1998meteor}. In contrast, \citet{Capek2019,Vojacek2019} found that many (though not all) irons that were identified by their spectra show early light curve peaks, and will thus have low F parameters. As a point of reference, 45 iron candidates identified by \citet{Capek2019} had an average F parameter of $0.31 \pm 0.04$. They note that some of their spectrally identified irons have higher F parameters.

In addition to early peaking lightcurves,  \citet {Capek2019,Vojacek2019} found that irons also tend to have short luminous trajectories compared to meteors of similar total integrated brightness/mass. These short trail lengths are a natural outcome of the proposed ablation model of \citet{Capek2019} whereby the rapid removal of liquid iron from ablating iron meteoroids more quickly removes mass as compared to normal thermal ablation through vaporization. This melting as a primary mode of ablation is very efficient at quickly removing material giving rise to both low F parameters and short luminous trajectories.

From the foregoing results of the work of \citet {Capek2019,Vojacek2019}, we distinguish iron meteoroids from other populations in our video data by selecting events with F parameters below 0.31, and trajectories shorter than 9 km\textemdash both are the average value for iron candidates in \citet{Capek2019}. The 9 km limiting length could have been set higher; for the lowest begin heights, the F parameter does not strongly change with trajectory length (see Section \ref{sec:results}). Both discriminators have exceptions; it is possible for irons to have higher F parameters and longer trajectories. But by choosing conservative cutoffs and imposing multiple conditions we can have high confidence that our selected iron candidates are in fact iron meteors, though our resulting numbers are likely lower limits.

To further strengthen our selection process, there is one more iron discriminator used in this study. As the heat of vaporization of iron is significantly higher than its melting point, an iron must first melt to begin spraying droplets. Additionally, when an iron meteoroid is heated, a significant amount of heat energy is conducted inwards. More energy is therefore needed to bring the surface layer of the meteoroid to its melting temperature \citep{Vojacek2019}. 

To acquire this extra energy, irons must penetrate deeper into the atmosphere as compared to regular stony meteoroids. The energy ($E_\mathrm{S}$) accumulated by a single body meteoroid from collisions with atmospheric molecules per cross-sectional area is given by:
\begin{align}
    E_\mathrm{S} = \frac{1}{2} \Lambda \frac{v^2}{\sin \theta} \int_{h_0}^{\infty}\mathrm{\rho}(h)\mathrm{d}h
\end{align}
Where $v$ is the initial velocity (which we approximate as constant over the luminous trajectory), $\rho$ is the atmospheric density as a function of height $h$, $\theta$ is the elevation angle from horizontal, and the integral is from the begin height $h_0$ to infinity \citep{borovivcka2007atmospheric}. We assume the heat transfer coefficient $\Lambda$ from the air stream is unity, a reasonable assumption for meteoroids in free-molecular flow conditions appropriate to our data \citep{popova2019}. 

\begin{figure}
    \centering
    \includegraphics[width=\linewidth]{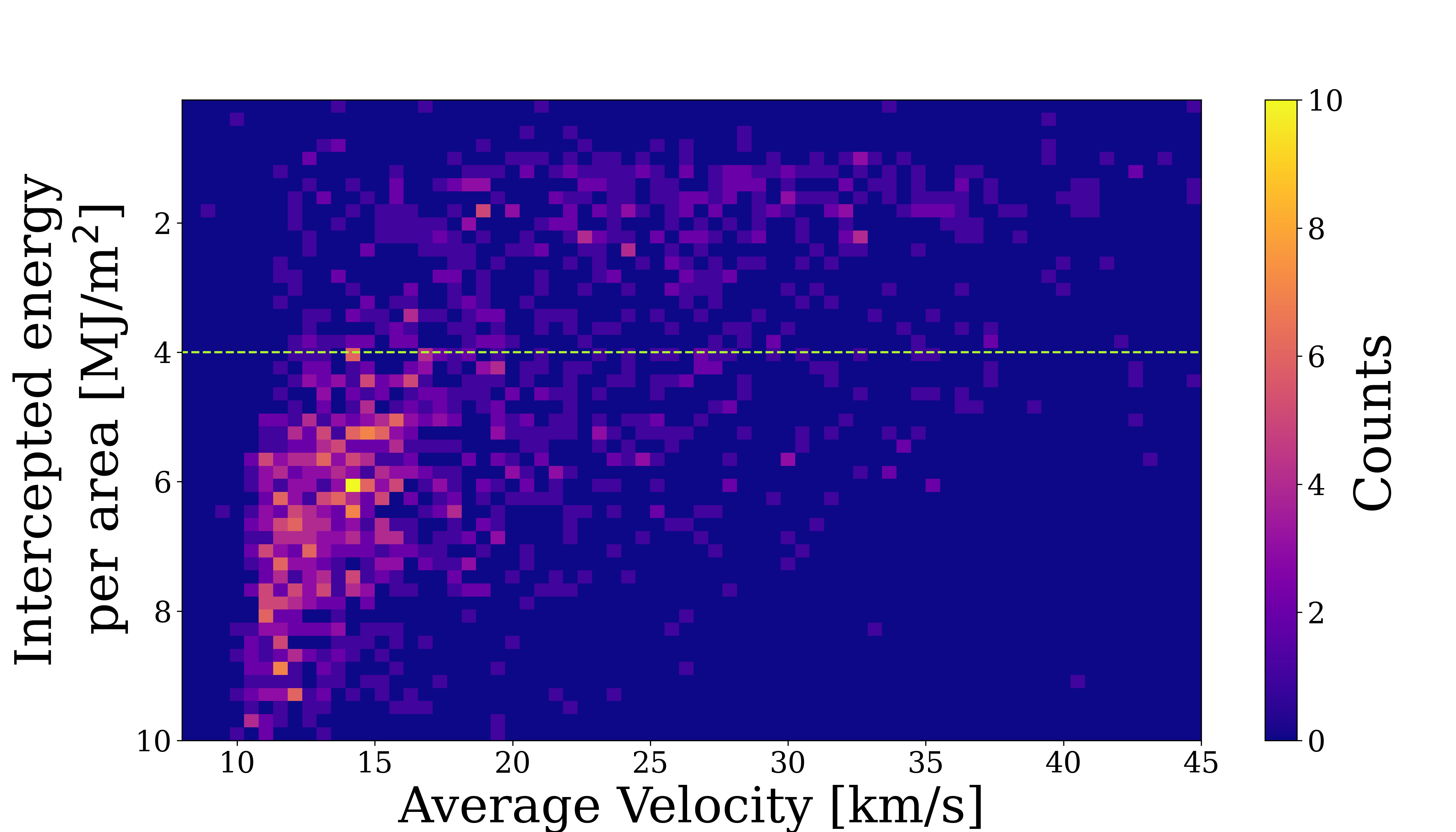}
    \vspace{-0.5cm}
    \caption{Density of iron-like meteors, binned by accumulated energy per area and velocity. Plotted meteors have F parameters of less than 0.31 and luminous trajectories shorter than 9 km. An energy per area line of $E_\mathrm{S}=4$ MJm$^{-2}$ for vertical entry is plotted in green, under which are the majority of iron candidates meeting our trail length and F parameter criteria.} 
    \label{fig:iron_fraction_contoured}
\end{figure}

Figure \ref{fig:iron_fraction_contoured} shows the number density of iron-like meteors binned by speed and accumulated energy that are isolated as likely irons utilizing the F parameter and trajectory length cutoffs previously described. Also shown is an energy per area isocurve corresponding to $E_\mathrm{S}=4$ MJm$^{-2}$ for vertical entry. \citet{Vojacek2019} found that the energy per unit area needed to begin spraying droplets for iron meteoroids was $\approx$ 8 MJm$^{-2}$ (see their Figure 7). Because irons also conduct heat inward, and the ratio of volume to cross-sectional area scales with radius, we expect our smaller-volume irons to have lower accumulated energies per area.

The value 4 MJm$^{-2}$ was chosen such that all iron candidates identified in \citet{Vojacek2019} and only the low-beginning meteors of Figure \ref{fig:iron_fraction_contoured} had higher accumulated energies per unit area. Moreover, the only significant number of non-iron meteors with higher accumulated energies in \citet{Vojacek2019} than this value were on sun-approaching orbits (which we find almost none of our iron events displayed as discussed later), and Halley-type events that constitute a small fraction of our final iron population (with none at low speeds where our iron population dominates).

Note that we incorporate the entry angle per event in computing this value (so the line above may be slightly lower for lower entry angle events). We require that our iron meteoroids must begin at lower heights; thus this $E_\mathrm{S}$ constraint is another discriminator to isolate possible iron meteors because the energy required for them to ablate is so high. 

We note that the only other spectral class of meteoroids which have similarly high energy per unit area interception values for onset of fragmentation/erosion are the sodium-free meteoroids \citep{Borovicka2005, Vojacek2019}. This implies there may be some contamination when selecting irons from this population using our criteria. Fortunately, the Na-free population has distinct orbital characteristics \citep{Vojacek2019, Matlovic2019a}, being in almost exclusively sun-approaching orbits ($q<0.2$ AU), intermediate perihelia ($q<0.4$ AU) for larger cm-sized irons, or Halley-type orbits \citep[e.g. alpha Monocerotids;][]{Borovicka2005}. This may be used to further discriminate between the iron and Na-free populations and is further discussed in Section \ref{sec:orbital_characteristics}. 

While previous studies have noted that meteors classified spectrally as irons tend to be in asteroidal orbits and have low entry velocities, we do not use these properties as iron discriminators, but restrict ourselves to physical attributes specific to the ablation process. While asteroidal origins are the most obvious parent candidates, the possibility still exists for iron meteoroids to come from other sources, possibly on highly evolved orbits. For example, \citet{Vojacek2019} noted two irons on Halley-type orbits. 

From all the foregoing, we argue that regardless of orbit, the above discriminators\textemdash low F parameter, short luminous trajectory, and high accumulated energies per area\textemdash are reasonable physical proxies of ablation to distinguish iron-rich meteoroids from cometary and stony material. Thus we aim to correlate the orbital origins of small iron meteoroids using these metrics.

\section{Results}
\label{sec:results}

\begin{figure}
    \centering
    \includegraphics[width=\linewidth]{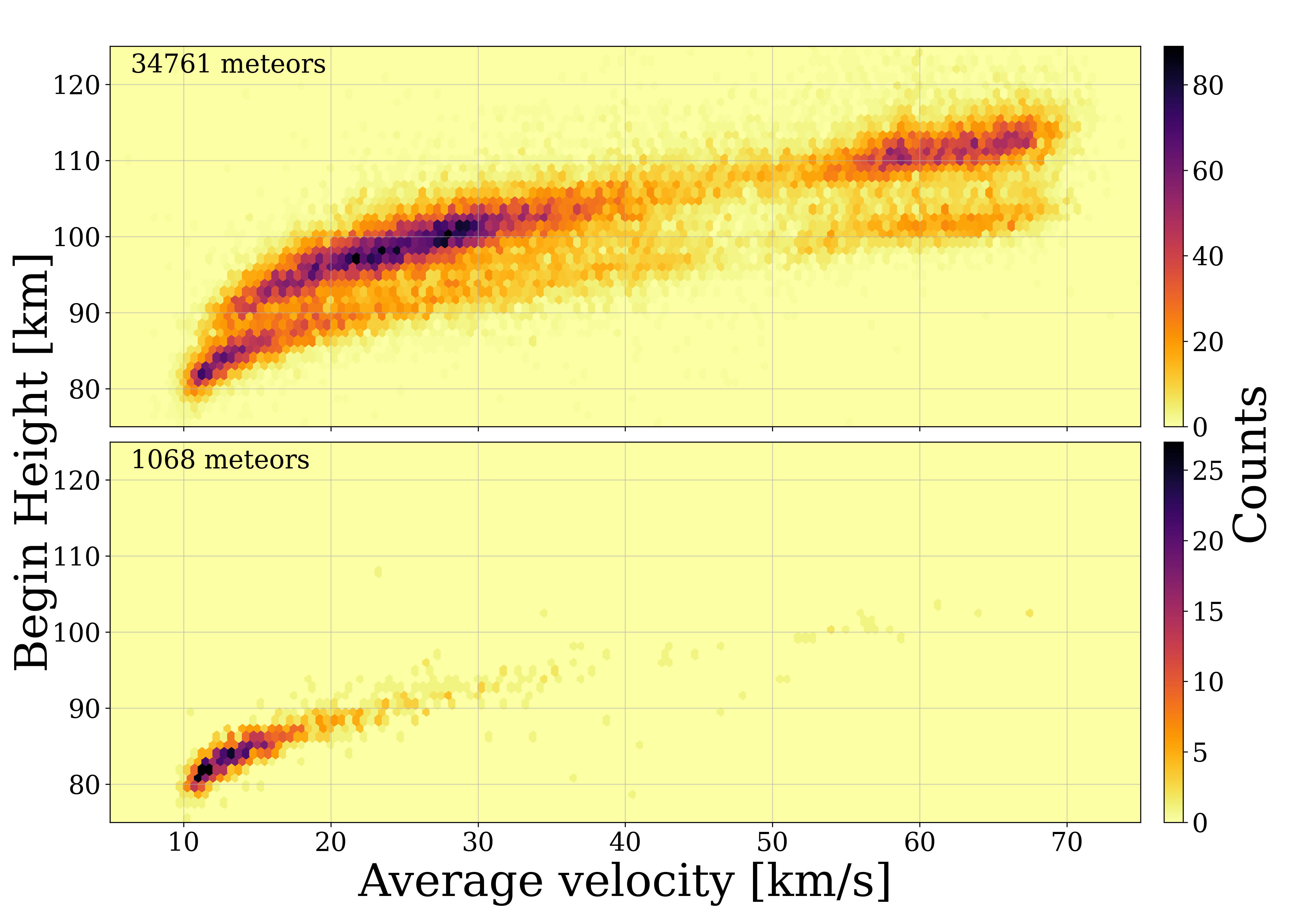}
    \vspace{-0.5cm}
    \caption{Density plot of begin height as a function of velocity for all meteors (top panel) and iron candidates (bottom panel).}
    \label{fig:ht-density}
\end{figure}

\begin{figure}
    \centering
    \includegraphics[width=\linewidth]{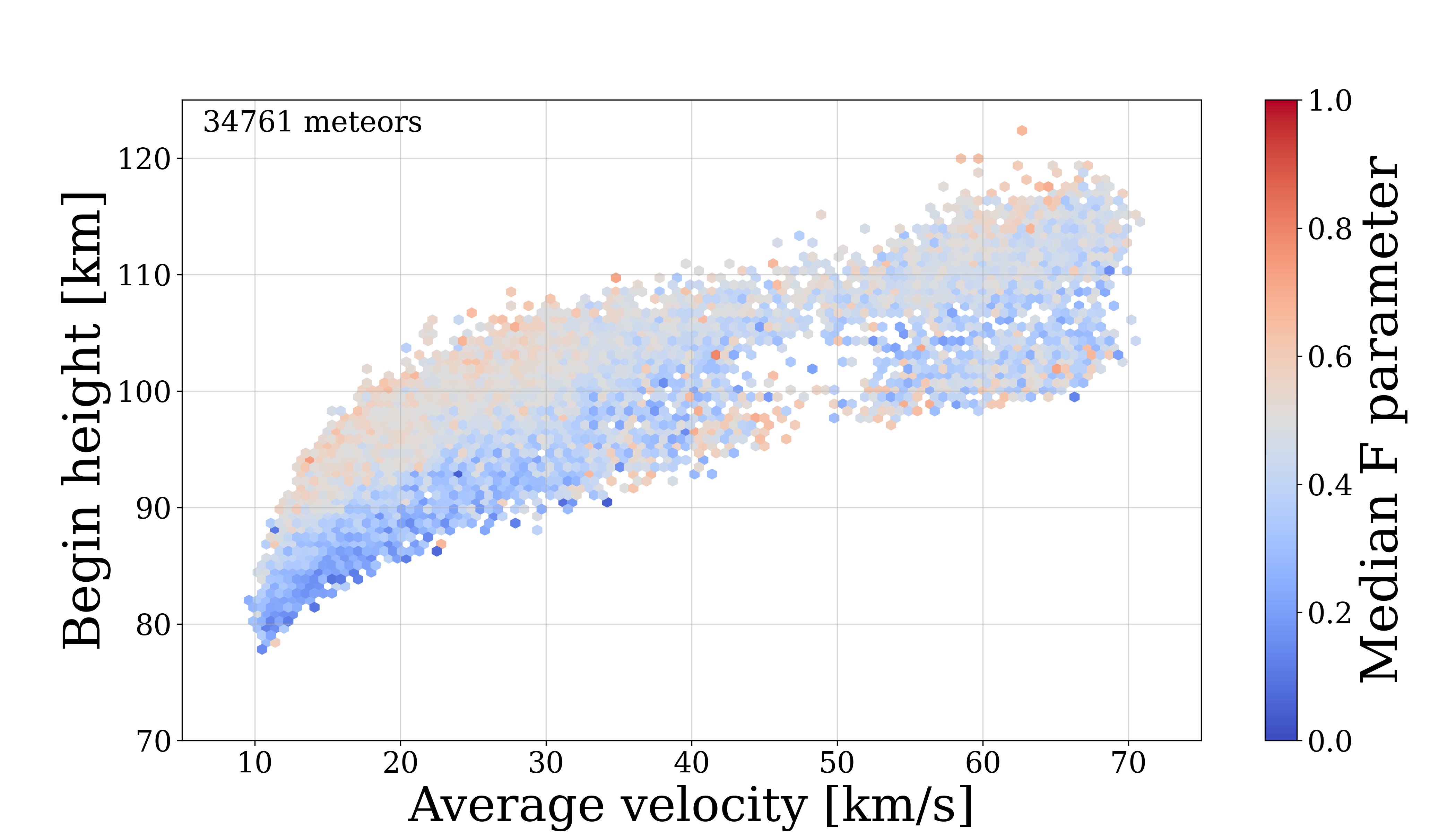}
    \vspace{-0.5cm}
    \caption{Median binned F parameter across different begin heights and average velocities. All meteors must have their beginning and ending observed in at least one camera, and three meteors must be present in a bin to be plotted.}
    \label{fig:f_parameter}
\end{figure}

In Figure \ref{fig:ht-density} we show the relative number density of meteor begin heights, for all meteors and our iron candidates, as a function of velocity. The most obvious feature is the well-known bifurcation in the meteoroid population \citep{Ceplecha1988} between cometary-type material (higher band of begin heights) and stronger, asteroidal-type material (lower band of heights).  

We begin by examining the trend in F parameter with begin height among these populations as a function of speed for our entire dataset of 34761 meteors with complete lightcurves as shown in Figure \ref{fig:f_parameter}. Here each symbol represents the median F value for all meteors falling in a particular hexbin. 

\begin{figure}
    \centering
    \includegraphics[width=\linewidth]{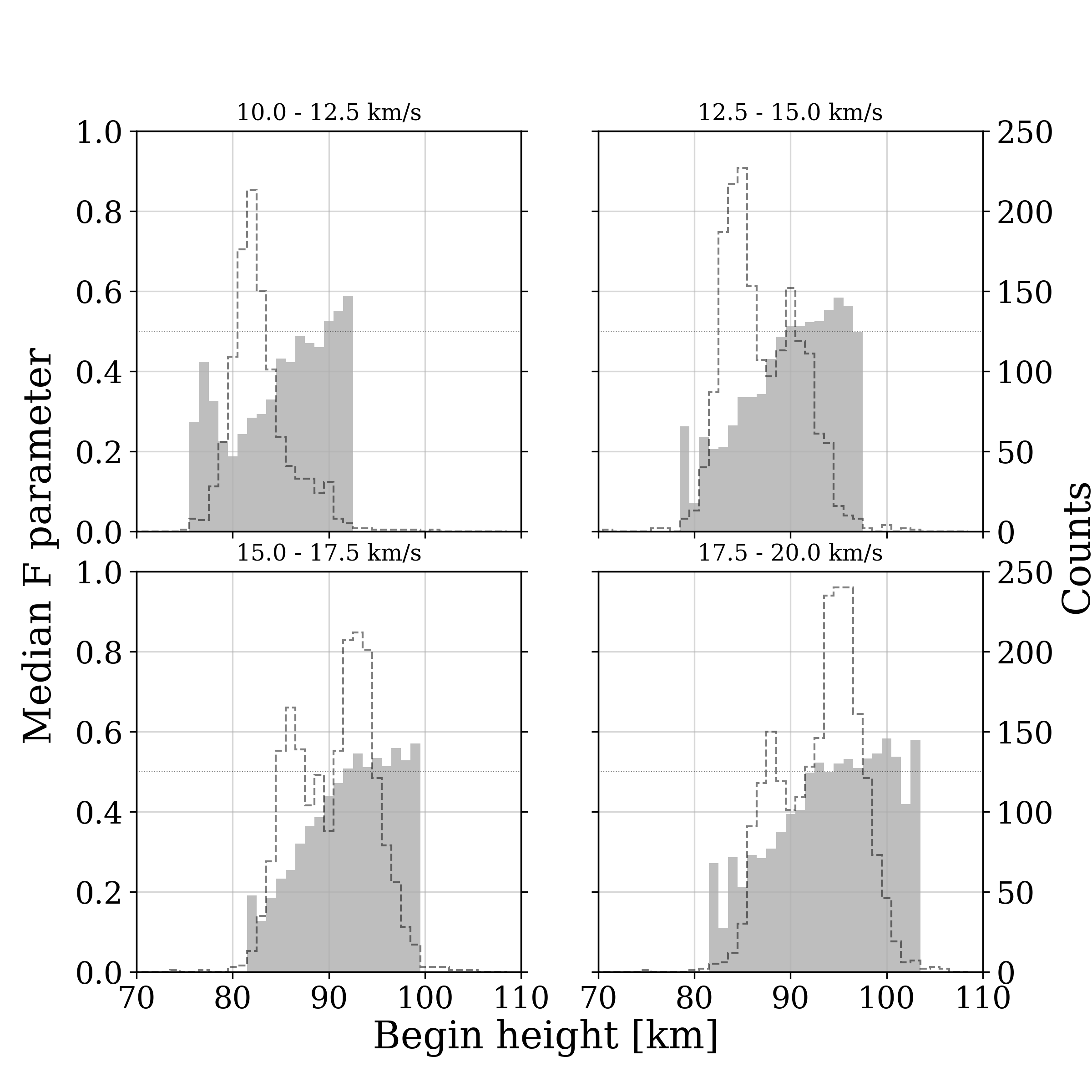}
    \vspace{-0.5cm}
    \caption{Median binned F parameter (left axis) across different begin heights (vertical bars) for the four lowest velocity bins. All meteors must have their beginning and ending observed in at least one camera. The number of meteors in each bin is shown by the dotted line and the right hand axis.}
    \label{fig:f_parameter_across_height}
\end{figure}

There is a gradient of decreasing F parameter with decreasing begin height, seen most strongly at low speeds. From prior work \citep[e.g.][]{Jacchia1958, Ceplecha1968, Jones1985, Ceplecha1988} begin height of ablation is a strong proxy for distinguishing between physically distinct meteoroid populations. A priori, we expect at the higher begin heights the lower density, weaker/lower material strength, and dustball type meteors to begin ablating due to the lower energy needed for the onset of ablation \citep{Ceplecha1958}. Meteors with this type of structure may release their grains before significant thermal ablation occurs and would have F parameters of around 0.5 \citep{Hawkes1975}. 

In contrast, refractory material, which need higher temperatures to commence ablation, begin at lower heights. We might expect this material to possess larger F parameters as they will ablate more nearly as single body theory predicts, with an F parameter close to 0.7 \citep{Beech2009}. However, we find a strong trend of low F parameters at low begin heights, a result similar to \citet{Campbell-Brown2015}. This indicates that the low, slow population of meteors ablates quickly and shows sudden onset in luminosity. Together with the low begin height, this is suggestive of a population dominated by iron-rich material ablating through melting, rather than refractory stony material, as first suggested by \citet{Borovicka2005}.  

In Figure \ref{fig:f_parameter_across_height} we see trends in the median F parameter as a function of begin height in fixed entry velocity windows. We see a clear tendency for F to be low at lowest heights and then reach a constant value near 0.5 above 90 km. We interpret this to reflect the transition to the higher, weaker dustball population where early fragmentation dominates the lightcurve shape. 

An interesting feature in Figure \ref{fig:f_parameter_across_height}, most visible in the lowest speed bin, is an apparent increase in median F at the very lowest begin heights. In general these still fall below our F cutoff of 0.31, but the local peak could be an indication of a mixed population of iron meteoroids with a non-iron refractory population ablating more as single bodies, driving the median F higher. It could also reflect differences in iron ablation behaviour with height. A weaker similar feature is visible in the higher speed bins, but small number statistics preclude any clear interpretation. 

One obvious means of creating a small F parameter is through sudden release of grains. However, it is difficult to reconcile such a process with the low begin heights we observe for the dominant low-F population; catastrophic grain release is associated more generally with weaker material. An alternative explanation is that these represent iron meteors. Iron meteors only become visible after melting \citep{Capek2019} and also begin deep in the atmosphere \citep{Vojacek2019} and are therefore more consistent with the observed behaviour of the low, slow population. 

\begin{figure}
    \centering
    \includegraphics[width=\linewidth]{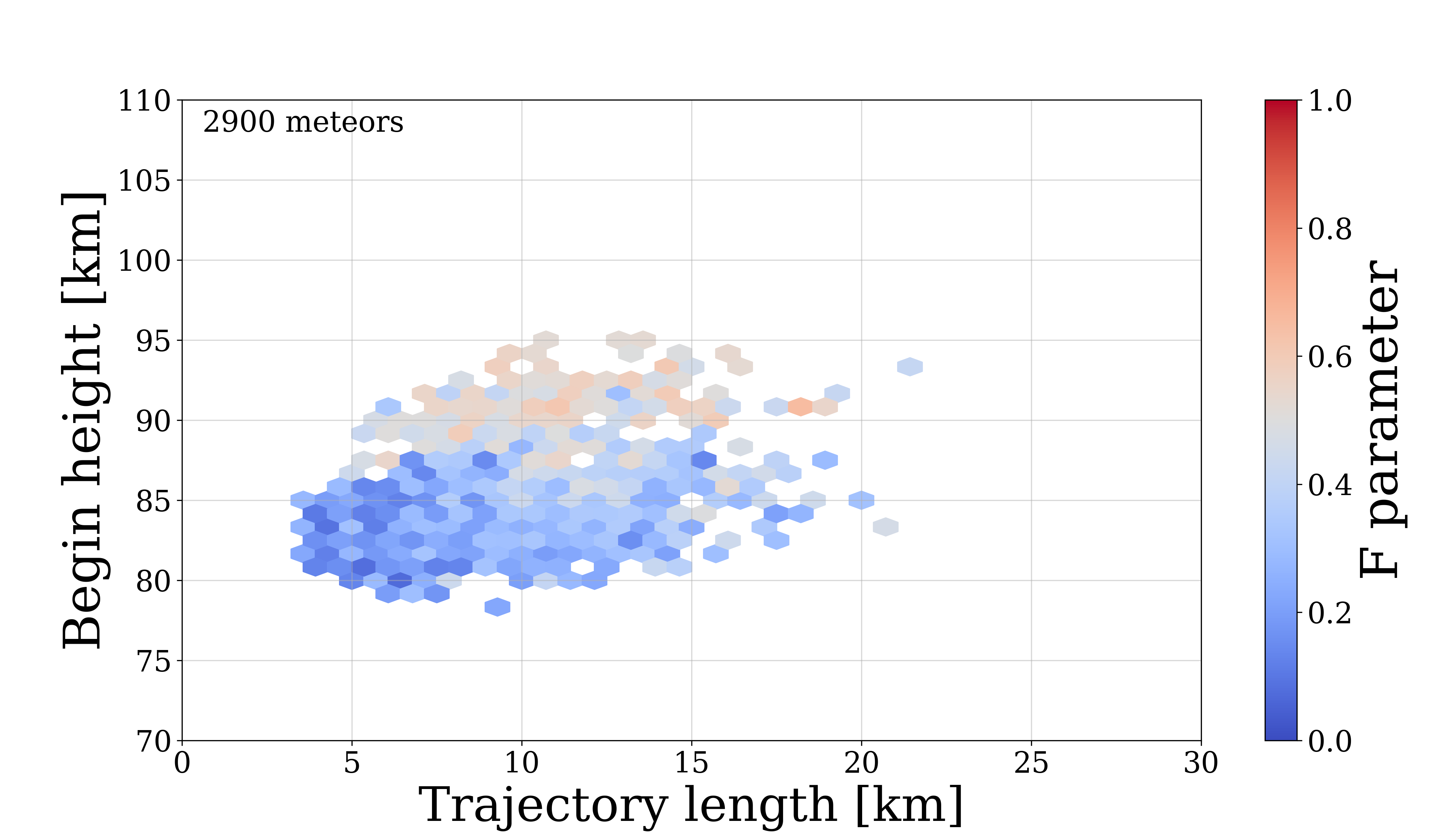}
    \vspace{-0.5cm}
    \caption{Begin height as a function of total path length showing median F parameter for meteors with measured speeds from 10-15 km/s.}
    \label{fig:ht-f-length}
\end{figure}

That some form of fragmentation is causing the low-F parameter is further supported by the apparent trail lengths of meteors as a function of F as shown in Figure \ref{fig:ht-f-length}. This figure shows only meteors in the 10-15 km/s velocity bin, where the low-slow population is most numerous, but the trend is similar in higher velocity bins. There is a noticeable correlation between low begin heights, short trail lengths and the smallest F-parameters. This suggest that meteors skewing to early brightness peaks have shorter trails lengths, as expected for fragmenting meteoroids. The trend in the figure also supports our adoption of 9 km limiting trail lengths, as the median F remains low even at longer trail lengths. This is consistent with the behaviour expected for iron meteors, though the fragmentation would in this case be in the form of iron droplets \citep{Capek2019}.

Using our criteria adopted earlier to identify probable irons, we select only meteors with  $F<0.31$, luminous trajectory lengths $l<9$ km, and energies accumulated per unit cross-sectional area before the beginning $E_\mathrm{S} > 4$ MJ. This produced 1068 total candidate iron meteors from among our total sample of 34761.

\begin{figure}
    \centering
    \includegraphics[width=\linewidth]{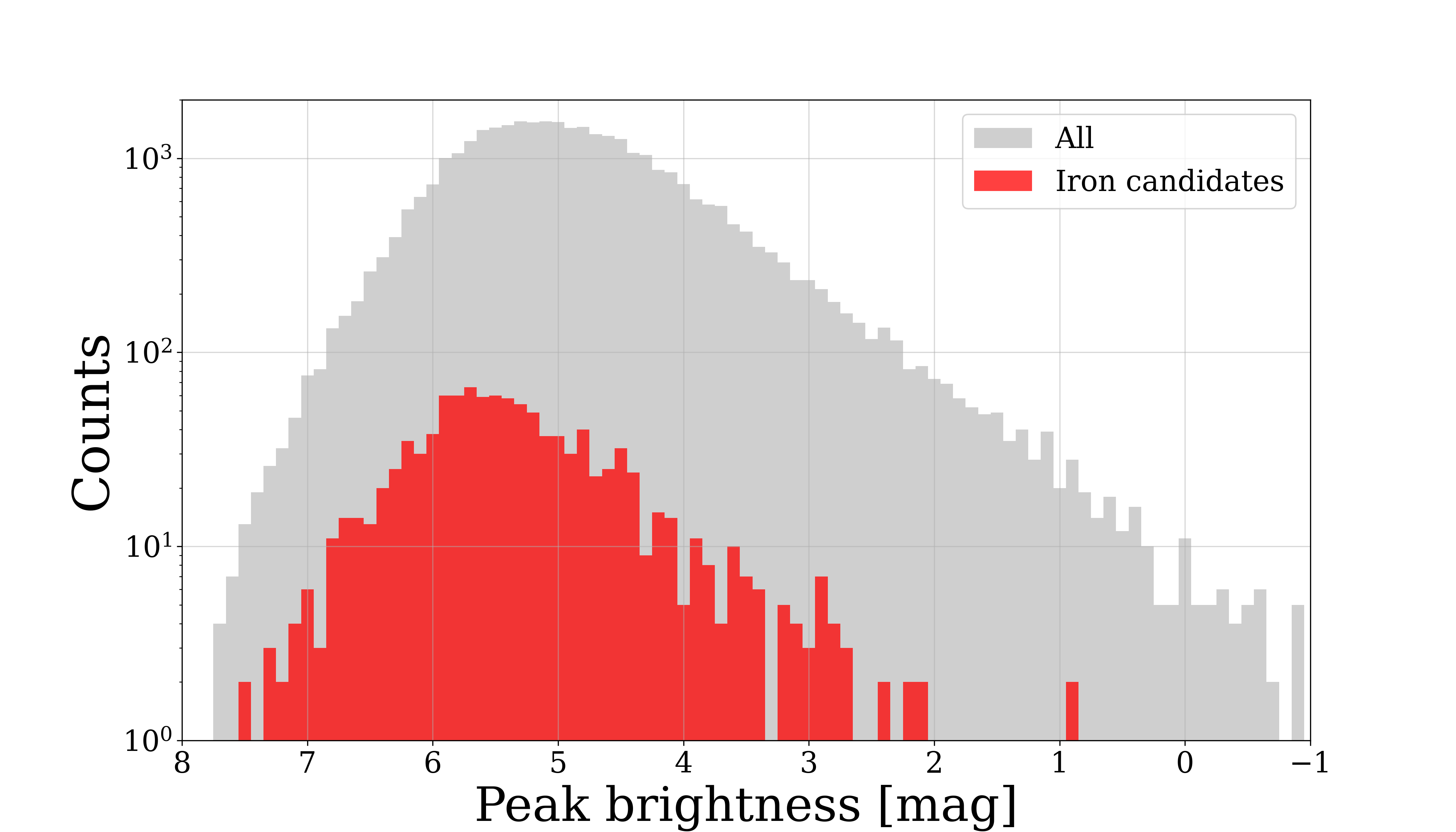}
    \vspace{-0.5cm}
    \caption{Distribution of peak magnitudes for iron (red) and the entire population as a whole. The effective survey completeness is to an absolute magnitude of +6 at low speeds using the methodology described in \citep{vida2020new}, falling to +5 at 60 km/s.}
    \label{fig:mag-dist}
\end{figure}

Figure \ref{fig:mag-dist} shows the distribution of peak magnitudes for the iron and entire population across all speeds. The overall survey is complete to an absolute magnitude of +6 at low speeds, falling to +5 at high speeds (60 km/s). This corresponds to masses of order 10$^{-5}$ kg at speeds of 15 km/s (assuming a fixed luminous efficiency of 0.7\%) and approaches 10$^{-7}$ kg at speeds of 60 km/s.

\begin{figure}
    \centering
    \includegraphics[width=\linewidth]{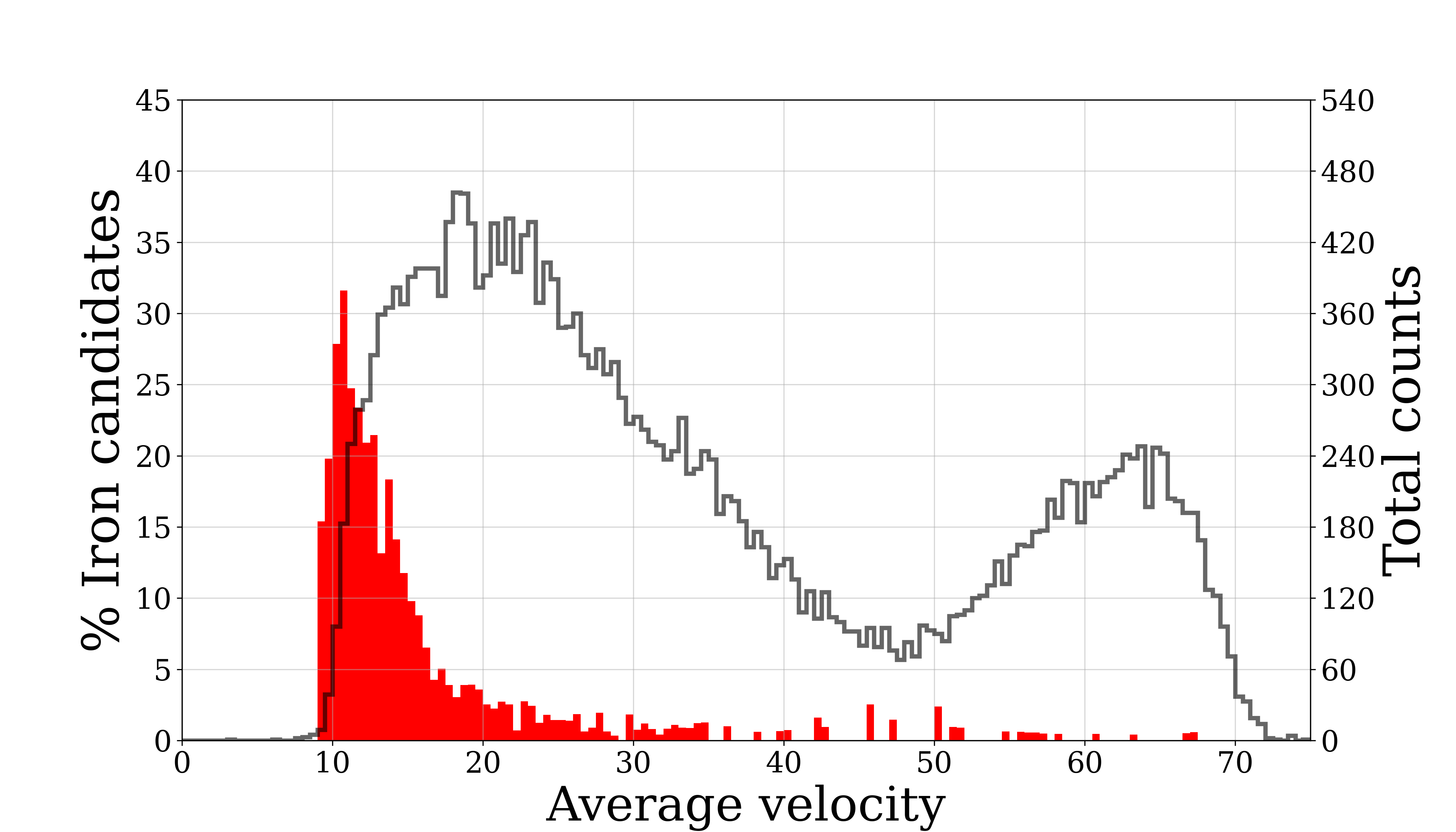}
    \vspace{-0.5cm}
    \caption{Fraction of the entire population (open dark histogram)  that are iron candidates (red vertical bars) at different speeds.}
    \label{fig:iron_fraction_hist}
\end{figure}

\begin{figure}
    \centering
    \includegraphics[width=\linewidth]{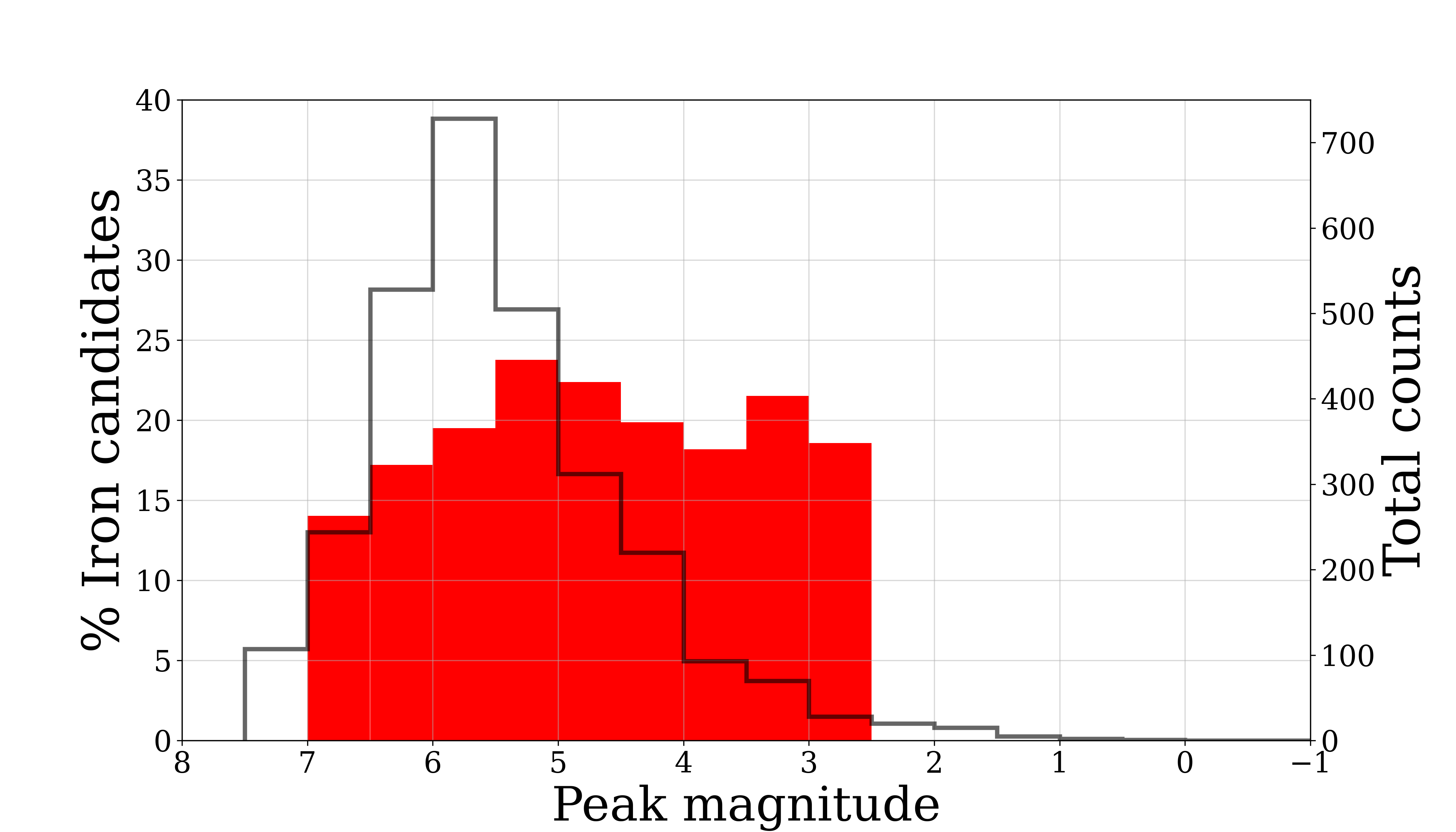}
    \vspace{-0.5cm}
    \caption{Raw, observed fraction (red vertical bars) of the entire population (open histogram) that are iron candidates at different peak absolute magnitudes in the 10-15 km/s velocity range. The total number of counts per bin is shown by the open histogram and right hand axis.}
    \label{fig:iron_fraction_mag}
\end{figure}

Figures \ref{fig:iron_fraction_hist} and \ref{fig:iron_fraction_mag} show the raw measured proportion of meteors that meet our criteria to be categorized as probable irons compared to the total population. At the lowest speeds, these iron candidates represent 20-25\% of all slow, mm-sized meteors between 10-15 km/s. At higher speeds ($>20$ km/s) the fractional values quickly fall to a few percent or less. There are a small number of potential iron candidates even at very high speeds. This may represent a small population of irons on Halley-type orbits as has been previously noted \citep{Vojacek2019} or may be contamination from the Na-Free population.

\begin{figure}
    \centering
    \includegraphics[width=\linewidth]{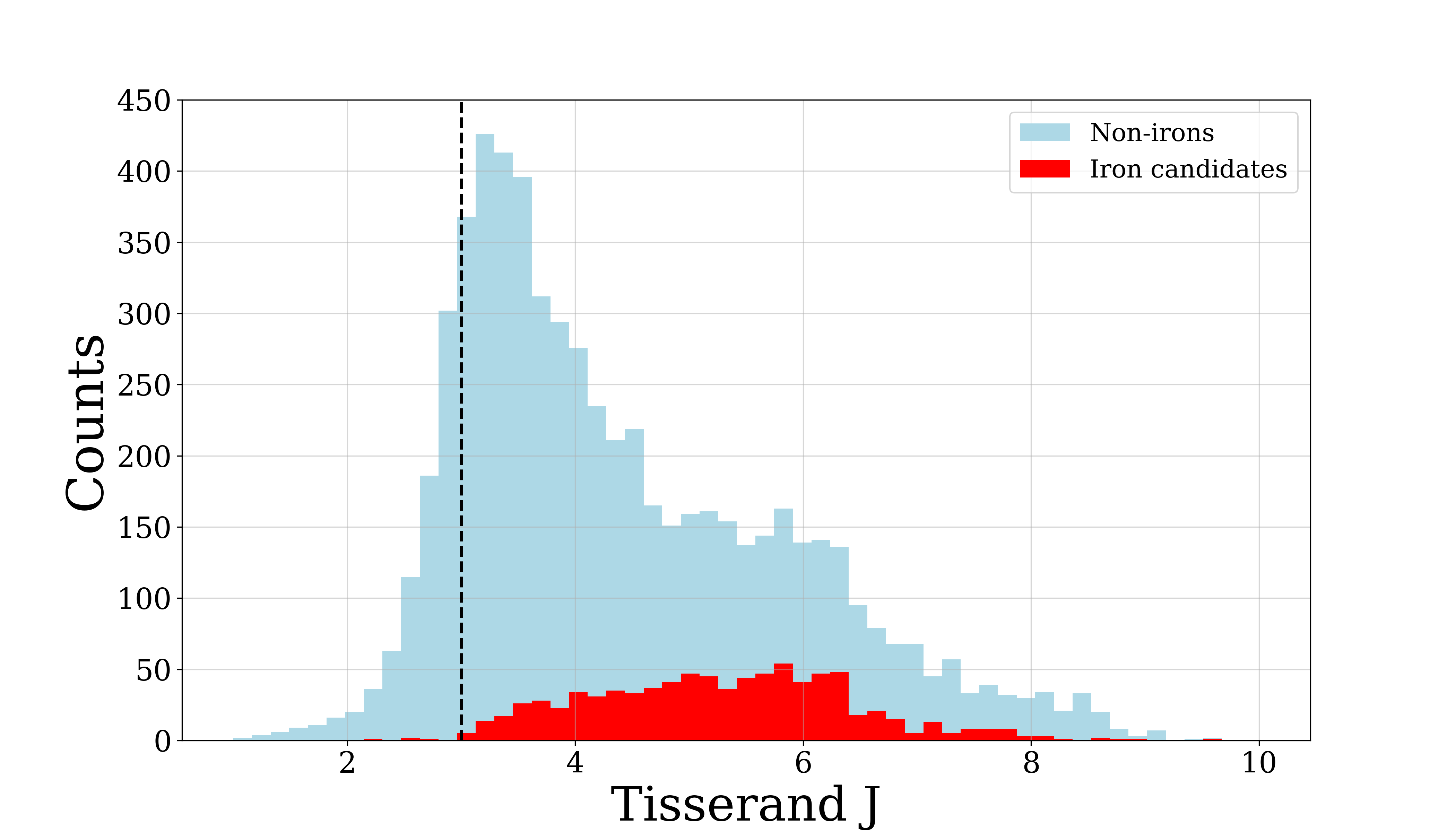}
    \vspace{-0.5cm}
    \caption{The Tisserand parameter with respect to Jupiter ($T_\mathrm{J}$) for iron candidates (red) and non-irons (blue) between 10-20 km/s. The iron population shows a significant drop across the $T_\mathrm{J}<3$ boundary (vertical dashed line).}
    \label{fig:iron_Tj_histo}
\end{figure}

\begin{figure}
    \centering
    \includegraphics[width=\linewidth]{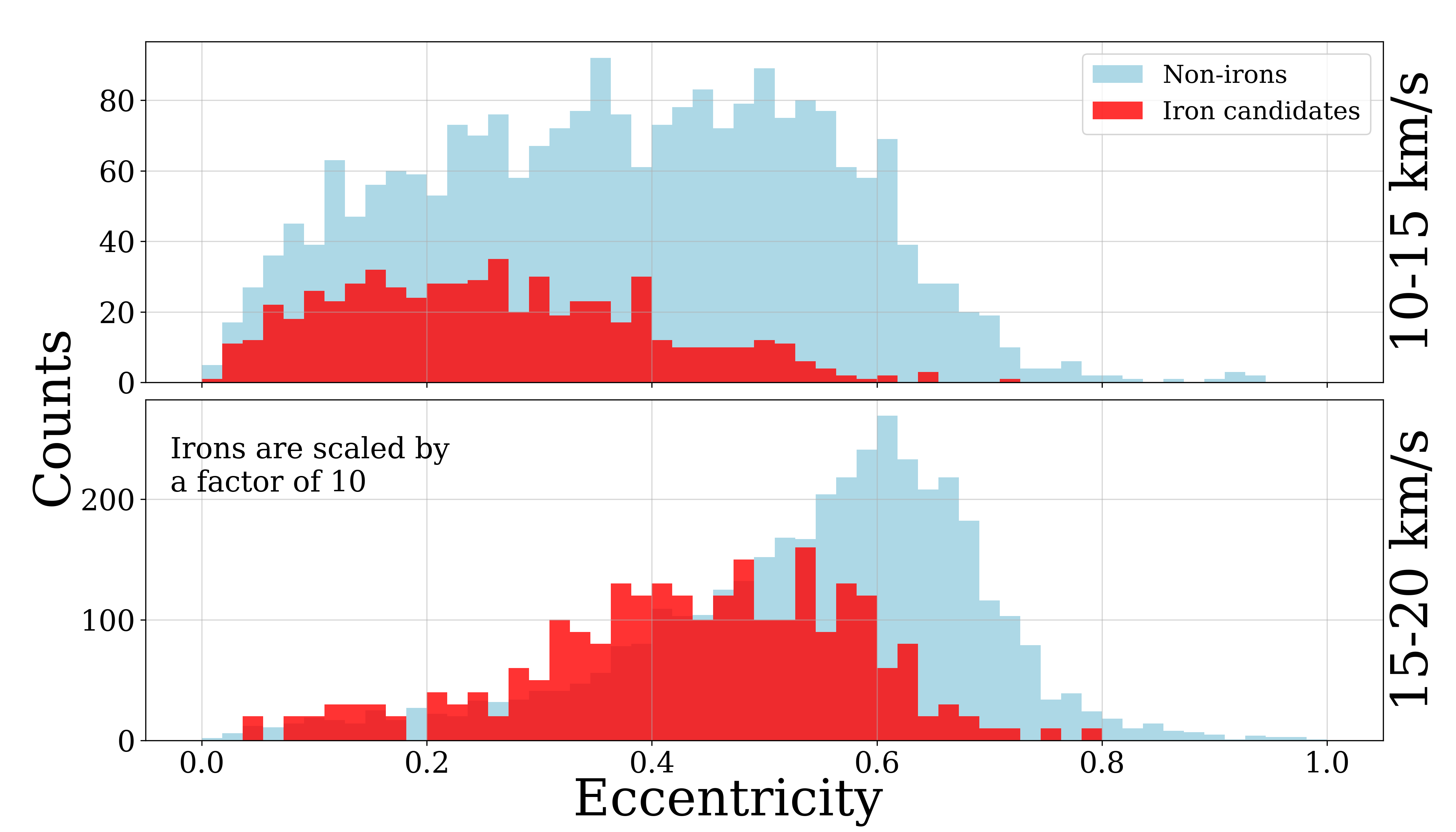}
    \vspace{-0.5cm}
    \caption{The eccentricity distribution of iron candidates (red) and non-irons (blue) between 10-15 km/s, and 15-20 km/s. Iron counts in the bottom panel have been scaled by a factor of 10 for easier visual comparison. Irons have systematically lower eccentricities than the rest of the population between 10-15 km/s.}
    \label{fig:e_hist}
\end{figure}


\begin{figure}
    \centering
    \includegraphics[width=\linewidth]{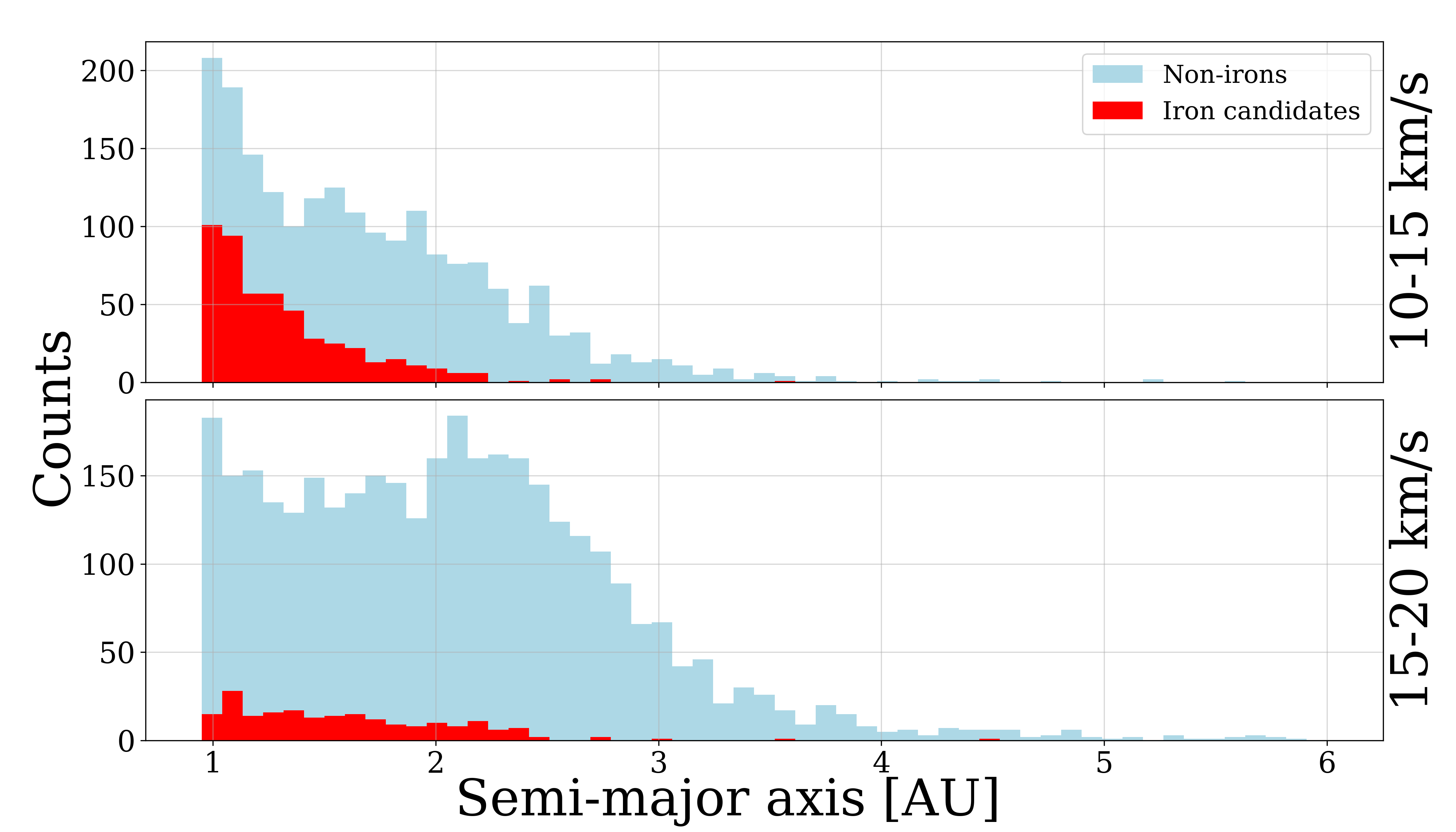}
    \vspace{-0.5cm}
    \caption{The semi-major axis distribution of iron candidates (red) and non-irons (blue) between 10-15 km/ and 15-20 km/s. Irons have systematically lower semi-major axis than the rest of the population.}
    \label{fig:a_hist}
\end{figure}




\begin{figure}
    \centering
    \includegraphics[width=\linewidth]{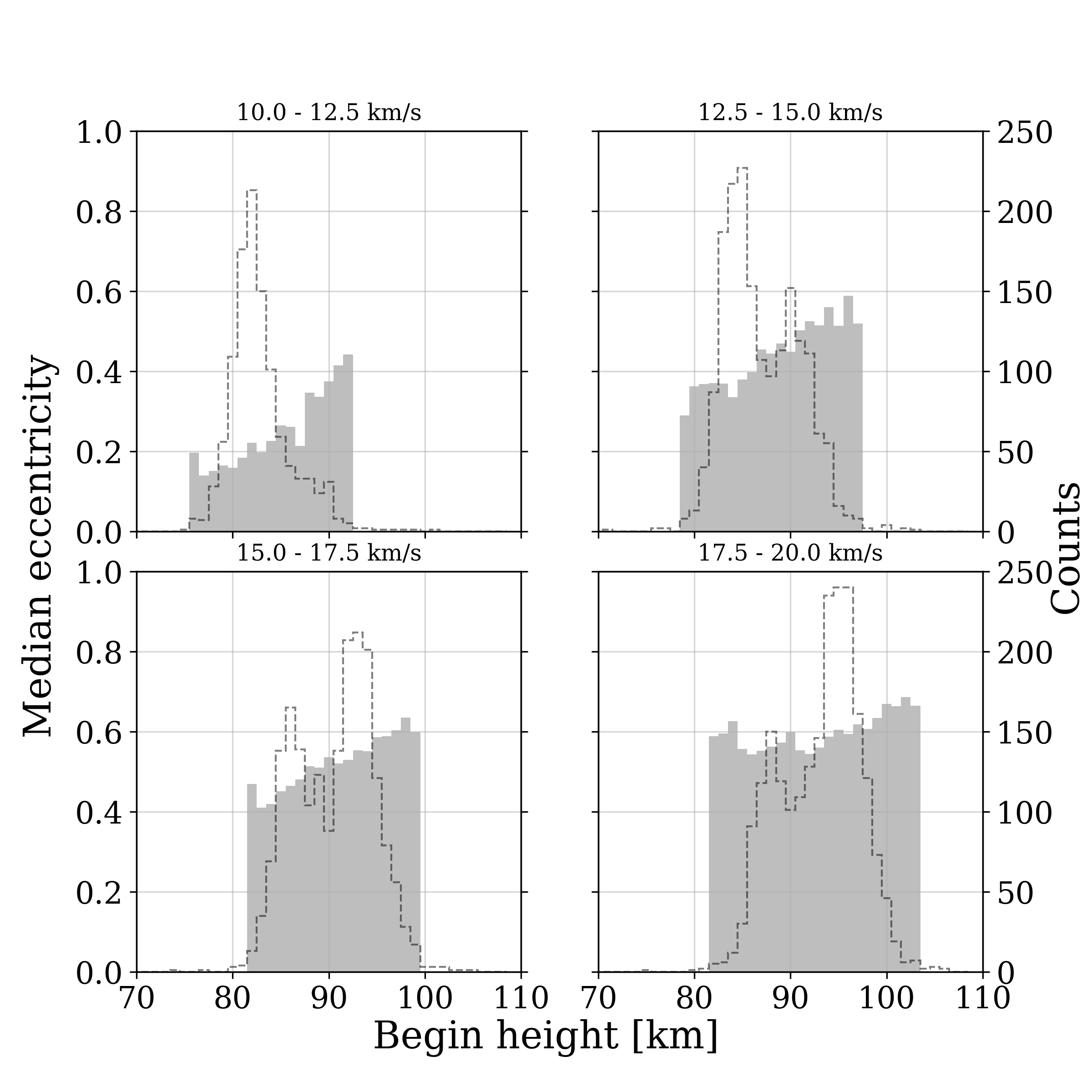}
    \vspace{-0.5cm}
    \caption{Median eccentricities of all meteors across begin height in slow velocity windows. At the slowest velocities where iron candidates are most abundant, the median eccentricity lowers.}
    \label{fig:e_hist_slow}
\end{figure}

\section{Discussion}
\label{sec:discussion}

\subsection{Orbital Characteristics}
\label{sec:orbital_characteristics}

The iron candidates chosen with the trajectory parameters outlined in Section \ref{sec:distinguishing} overwhelmingly possess asteroidal orbits.
Figure \ref{fig:iron_Tj_histo} shows the distribution of $T_\mathrm{J}$ for iron and non-iron meteoroids with entry speeds from 10-20 km/s, where irons are most abundant. As these cover the same range of velocities, there is no a priori reason the orbital distributions should differ for purely dynamical reasons. It is clear there is a strong cut-off in irons for $T_\mathrm{J}<3$, whereas a significant tail extends below 3 for the non-iron population. It is clear there is some JFC (and even HTC) material at these speeds in the non-iron population while essentially none in the iron population. Note that PR drag will tend to move objects from low $T$ to higher $T$ with time; hence the non-irons to the left of the $T_\mathrm{J}$ line are clearly from JFC parents and comparatively young. 

The inclinations of both populations in this velocity window are also low, with irons showing slightly lower inclinations than non-irons. A more dramatic difference is visible in the eccentricity and semi-major axis  distributions as shown in Figures \ref{fig:e_hist} and \ref{fig:a_hist}. Here we break down the distributions into 10-15 and 15-20 km/s bins to illustrate the large difference between the iron and non-iron populations, the former being on smaller, more circular orbits. 

This gradient in eccentricity is further evident in Figure \ref{fig:e_hist_slow} where the median eccentricity as a function of begin height in finer velocity intervals is shown. 

\begin{figure}
    \centering
    \includegraphics[width=\linewidth]{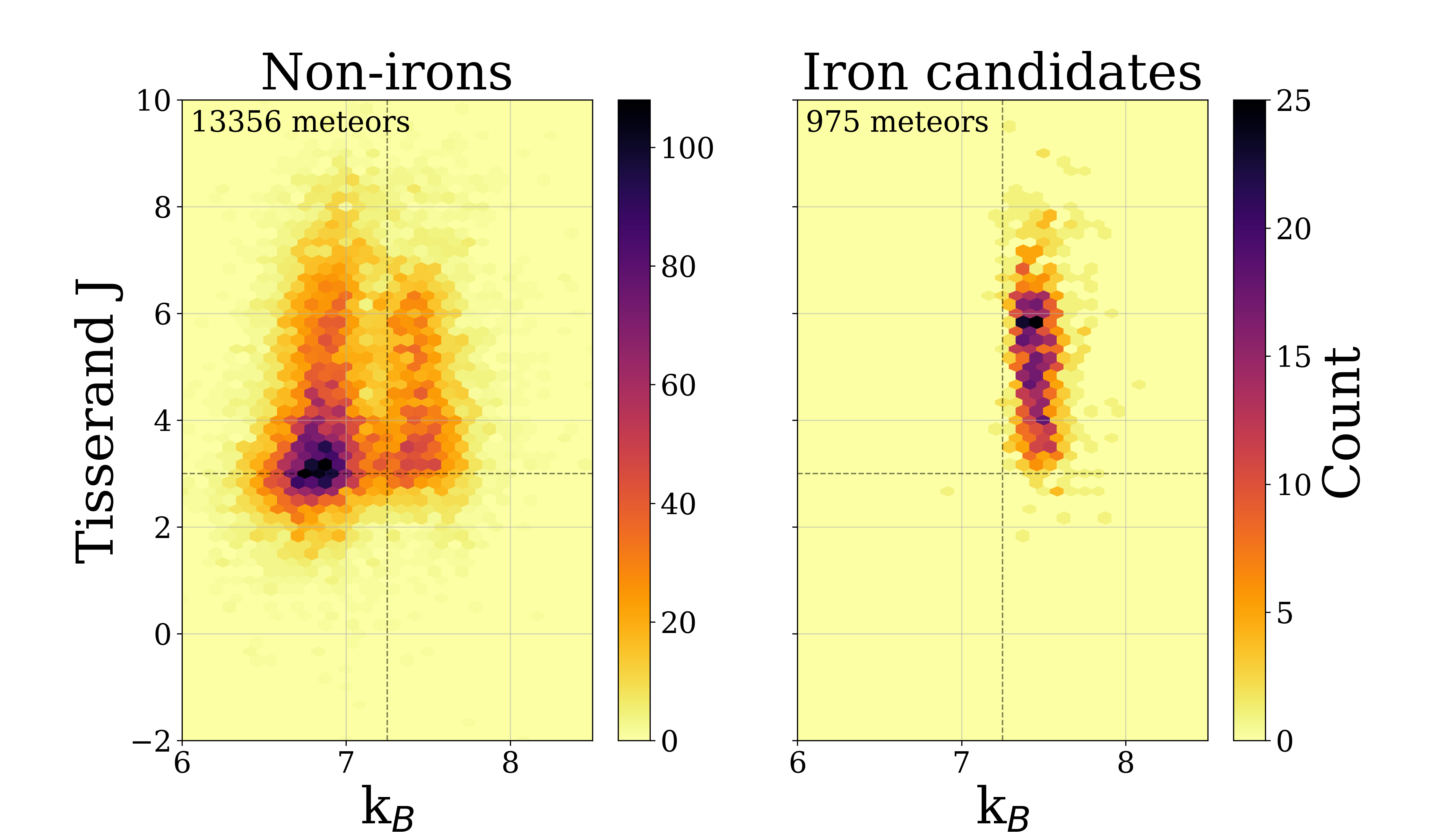}
    \vspace{-0.5cm}
    \caption{Tisserand parameter with respect to Jupiter versus the $k_\mathrm{B}$ parameter at velocities between 10 and 30 km/s. $k_\mathrm{B}=7.25$ and $T_\mathrm{J}=3$ marked to illustrate the cutoff of iron candidates, which are primarily asteroidal.}
    \label{fig:tj_kb_comparison}
\end{figure}

Figure \ref{fig:tj_kb_comparison} compares the $k_\mathrm{B}$ and $T_\mathrm{J}$ parameters of iron candidates and non-iron candidates with entry speeds between 10 to 30 km/s. The $k_\mathrm{B}$ parameter is a different measure of the energy required for ablation, and can be used to coarsely classify compositional populations of meteoroids \citep{ceplecha1967spectroscopic, Ceplecha1988}. It is given by:
\begin{align}
    k_B = \log\rho_{_{B}} + 2.5\log v_{\infty} - 0.5\log\cos z_{_{R}}
\end{align}
Where $\rho_B$ is the atmospheric density at the begin height, $v_\infty$ is the initial velocity and $z_R$ is the zenith angle. By enforcing high $E_\mathrm{S}$ values on our iron candidates they will naturally have high $k_\mathrm{B}$ parameters. The Tisserand parameter with respect to Jupiter on the other hand is independent and shows a stark cutoff at a value of 3. The iron candidates are dense, asteroidal material that are distinct from the blend of populations at the same velocities.

While the iron-candidates appear to be a distinct population of meteoroids, there is a possible degeneracy with Na-free and Na-poor type meteoroids. \citet{Borovicka2005}, and \citet{Vojacek2019,Vojacek2015} both note a population of refractory material that lacks Na lines in their meteor spectra. It is unclear whether this population would have short trajectories or low F parameters similar to irons, but they would have high $E_\mathrm{S}$ values. These Na-free and poor classes of meteoroids, however, can be broken down into two orbital sources: sun-approaching and cometary. Specifically, the vast majority of Na-free meteors have perihelion distances $q < 0.2$ AU. From Table \ref{tab:orbits_summary}, only a small portion of our iron candidates are on sun-approaching or Halley-type orbits. We conclude that there is not significant contamination of Na-free or Na-poor meteors in our selected iron population, which has dominantly asteoridal-orbit characteristics.

\begin{table}
	\centering
	\caption{Orbits of iron candidates and non-irons following the classification scheme of \citet{Borovicka2005}.}
	\label{tab:orbits_summary}
	\begin{tabular}{lcc}
	    \hline
	    Orbit & Iron candidates & Non-irons \\
	    Total counts & 1068 & 33641 \\
	    \hline
	    Sun-approaching & 1.7\% & 5.4\% \\
	    Ecliptic shower & 0\% & 0.5\% \\
	    Halley & 2.8\% & 38.7\% \\
	    Jupiter family & 0.7\% & 2.1\% \\
	    Asteroidal-chondritic & 96.2\% & 65.8\%
	\end{tabular}
\end{table}

\subsection{Possible Origin}

\begin{figure}
    \centering
    \includegraphics[width=\linewidth]{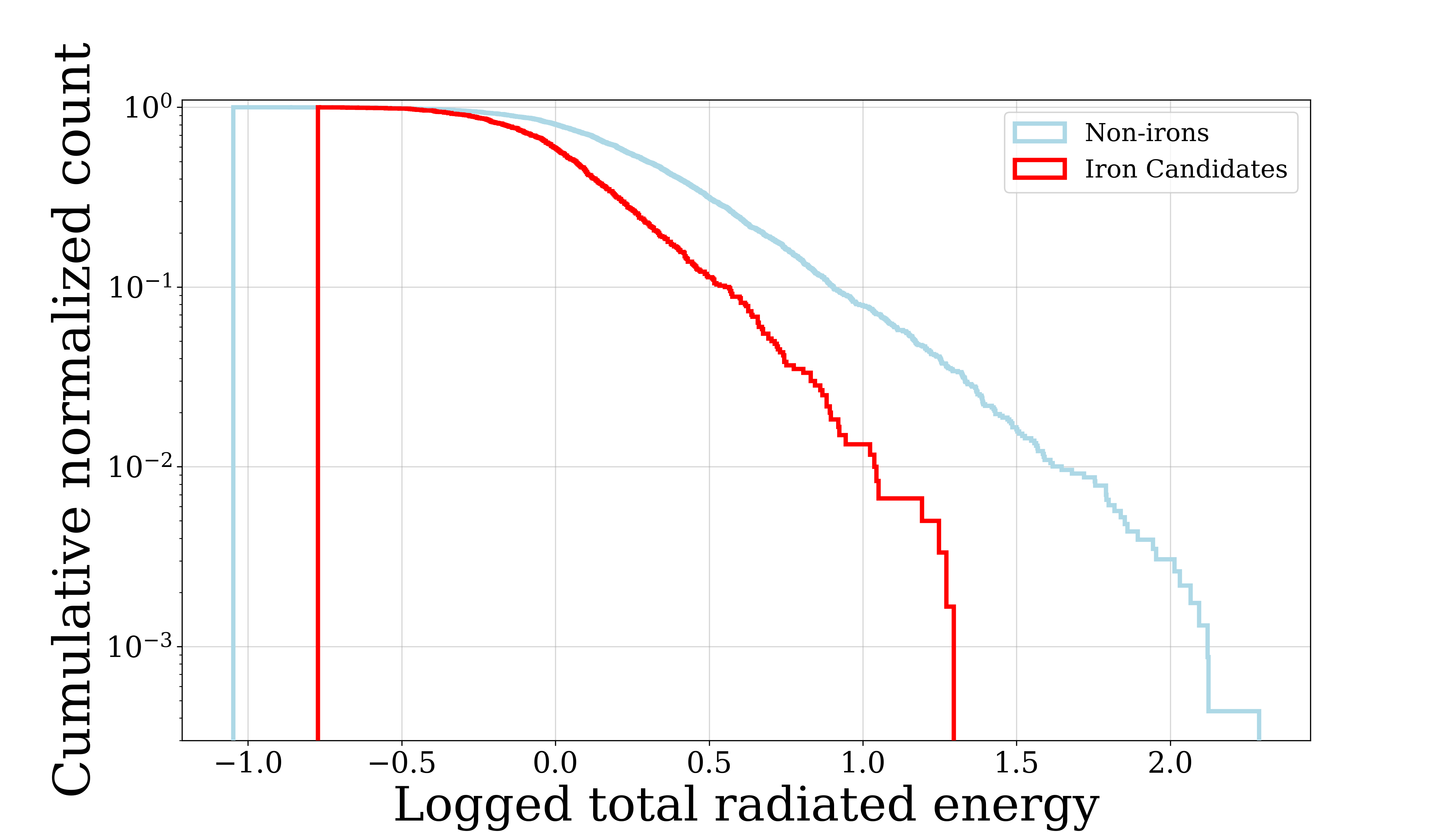}
    \vspace{-0.5cm}
    \caption{Cumulative, normalized histograms of iron candidates and non-irons at 10-15 km/s. Meteors with peak magnitudes brighter than +1 magnitude are excluded due to saturation effects. At the lowest speeds and higher energies, iron candidates grow in abundance with decreasing energy quicker than the general population.}
    \label{fig:cumulative_counts}
\end{figure}

Figure \ref{fig:cumulative_counts} shows the cumulative total radiated energy distributions of iron candidates and non-irons. The plot is limited to 10-15 km/s due to the small numbers of iron candidates at higher speeds. Additionally, meteors with peak brightnesses brighter than +1 mag are excluded due to saturation effects. The iron candidates have a relative deficit of high energy meteoroids compared to the general population at low speeds.

Metal rich chondrules (also called nodules) are found in some chondrites and range in reported sizes up to several mm \citep{Rubin1999}. These Fe-Ni rich chondrules, while rare in ordinary chondrites, are of order 10\% the total volume of EL Chondrites \citep{Friedrich2015}. We suggest that impact comminution of chondrites may free such metal nodules, which would have long collisional ages in analogy with iron meteorites that have an order of magnitude or more larger Cosmic Ray Exposure ages \citep{herzog2014cosmic} as compared to chondrites. Over time these main-belt derived impact products would then preferentially evolve to smaller, more circular Earth-crossing orbits through PR drag. 

We note that \citet{jenniskens2016cams} also recognized the unusual orbital characteristics of this slow, deeply penetrating meteoroid population. They ascribed the presence of these large (cm-sized), refractory meteoroids in small, lower eccentricity orbits as due to unusually long collisional lifetimes ($\approx$ 10$^6$ years). However, while the population they identify has similarities to our iron candidates, our iron candidate orbital distribution is less eccentric (Figures \ref{fig:e_hist} and \ref{fig:e_hist_slow}) with smaller semi-major axes (Figure \ref{fig:a_hist}) than the cm-sizes observed by CAMS. Moreover, our population shows an abrupt cutoff at $T_\mathrm{J}=3$ (see Fig \ref{fig:iron_Tj_histo}) compared to the non-iron meteoroid population at the same entry speeds, which suggests an asteroidal origin. 

While it is possible for JFC-released meteoroids to decouple from Jupiter under PR drag, the normal outcome for larger JFC meteoroids is ejection before PR effects allow dynamical decoupling \citep{Dikarev2002,Nesvorny2011} or collisonal disruption on much shorter timescales \citep{jenniskens2008meteoroid} before significant orbital circularization occurs. For the iron candidate meteoroids in our survey, sizes range from several mm up to a cm. The PR timescale for iron-objects to reach such low eccentricities from JFC-like orbits is of order 10$^{7}$ years.

For such large particles from the main-belt to reach the Earth purely through PR drag takes close to 10$^8$ years. This timescale is an extreme upper limit as objects in NEA-like Earth crossing orbits have dynamical lifetimes of <10$^7$ years. In reality the delivery process likely is similar to the delivery for near-Earth asteroids \citep{granvik2017escape} and meteorites \citep{morbidelli1998orbital, granvik2018identification}. In this picture, iron meteoroids would drift under PR drag to a resonance escape hatch in the main belt and then quickly (10$^6$ year timescales) be delivered to Earth crossing orbits through increases in eccentricity. They then evolve to smaller semi-major axis/eccentricites through continued PR drag. 

While of approximately the correct size and origin to explain the iron meteoroid population we observe, whether these nodules can be collisionally generated in sufficient abundance and survive both dynamically and collisionally for such long times to match the observed iron population remains an outstanding question. We leave the quantitative investigation needed to examine the veracity of this argument for future work.

\subsection{Biases}

Figure \ref{fig:height-overlap} shows the overlapping collection area of the two camera pairs used in this study. Since our meteors must be correlated between cameras, a counting bias is introduced; everything being equal we expect more meteors observed at heights with greater overlapping areas. Our iron candidates occur at lower heights than the majority of meteors and will thus be biased in number. To account for this bias, we find the overlapping area at each meteor's peak height for the camera pair it was observed in and assign it a weighting based on the reference height of 100 km. For detections that were seen in all four cameras, the camera pair with the greatest sum of frames is used. The overlapping area at 100 km for F cameras is 143.8 km$^{2}$, and for G cameras it is 786.1 km$^{2}$. 

\begin{figure}
    \centering
    \includegraphics[width=\linewidth]{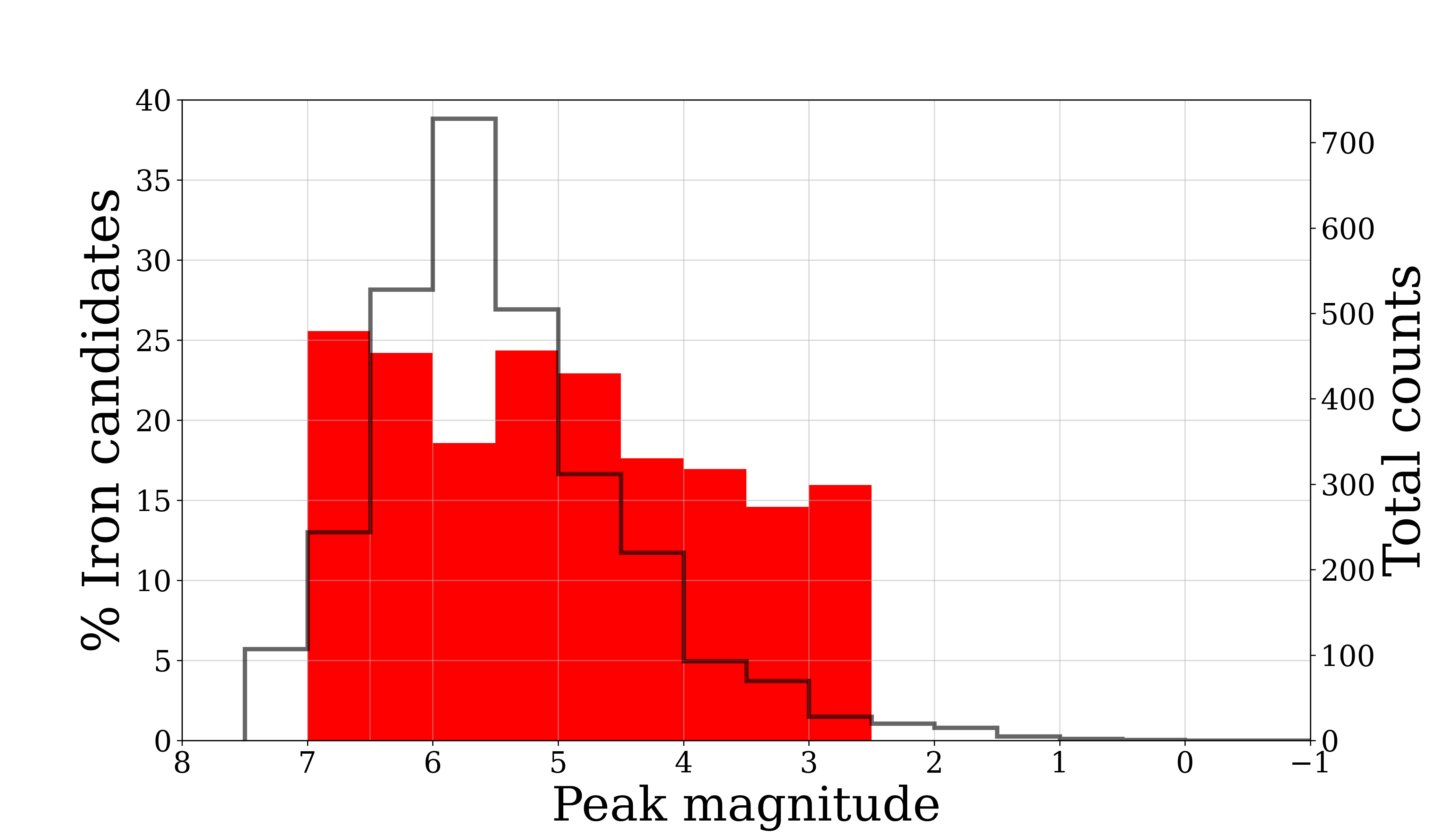}
    \vspace{-0.5cm}
    \caption{The de-biased fraction of iron candidates (red filled bars) as a function of magnitude (in 0.5 mag bins) with entry speeds between 10-15 km/s. The total number of events per 0.5 magnitude bin is shown on the right hand axis as the unfilled grey histogram.}
    \label{fig:mag_dist_debiased}
\end{figure}

In extreme cases, very small overlaps at low heights can cause weightings to be very high. Ignoring weightings with values greater than 10, the fraction of iron candidates at speeds between 10-15 km/s goes from 21.1\% to only 21.9\%. The number of meteors with weightings over 10 makes up 1\% of the sample. If the maximum weight of 100 is used instead, $<0.1\%$ of meteors are excluded and the fraction of iron candidates drops to 18.6\%. In this case, a few outliers impact the final weighted count, but the overall fraction of iron candidates still only changes by a few percent. However, the fraction as a function of magnitude changes more significantly. We settle on using a maximum weight of 10 for our final counts in Table \ref{tab:weighted_counts}. Figure \ref{fig:mag_dist_debiased} shows the trend in the de-based fraction of iron candidates as a function of absolute magnitude. 

\begin{table*}
	\centering
	\caption{Counts and percentages of iron candidates with range and overlapping area debiasing.}
	\label{tab:weighted_counts}
	\begin{tabular}{lccc}
	    \hline
	      & Iron candidates & All meteors & \% Iron candidate\\
	    \hline
	    Total counts & 1068 & 34761 & 3.0\% \\
	    Weighted total counts & 2444 & 49710 & 4.9\% \\
	    \hline
	    10-15 km/s counts & 600 & 2900 & 20.7\%\\
	    Weighted 10-15 km/s counts & 1489 & 6927 & 21.5\%
	\end{tabular}
\end{table*}

In addition to the overlapping area at different heights, the range can also bias our sample. The brightness of a meteor is proportional to the inverse of the range squared. Dim meteors far away from the cameras could therefore be lost. All meteor absolute magnitudes are calculated using a reference range of 100 km. Simply, meteors with ranges under 100 km will be more easily detected than meteors at ranges over 100 km. However, counter to this bias, only larger, brighter material will survive to low heights, also corresponding to low ranges. Hence at any given height we observe a mass filtered fraction of the total mass distribution impinging at the top of the atmosphere.

\begin{figure}
    \centering
    \includegraphics[width=\linewidth]{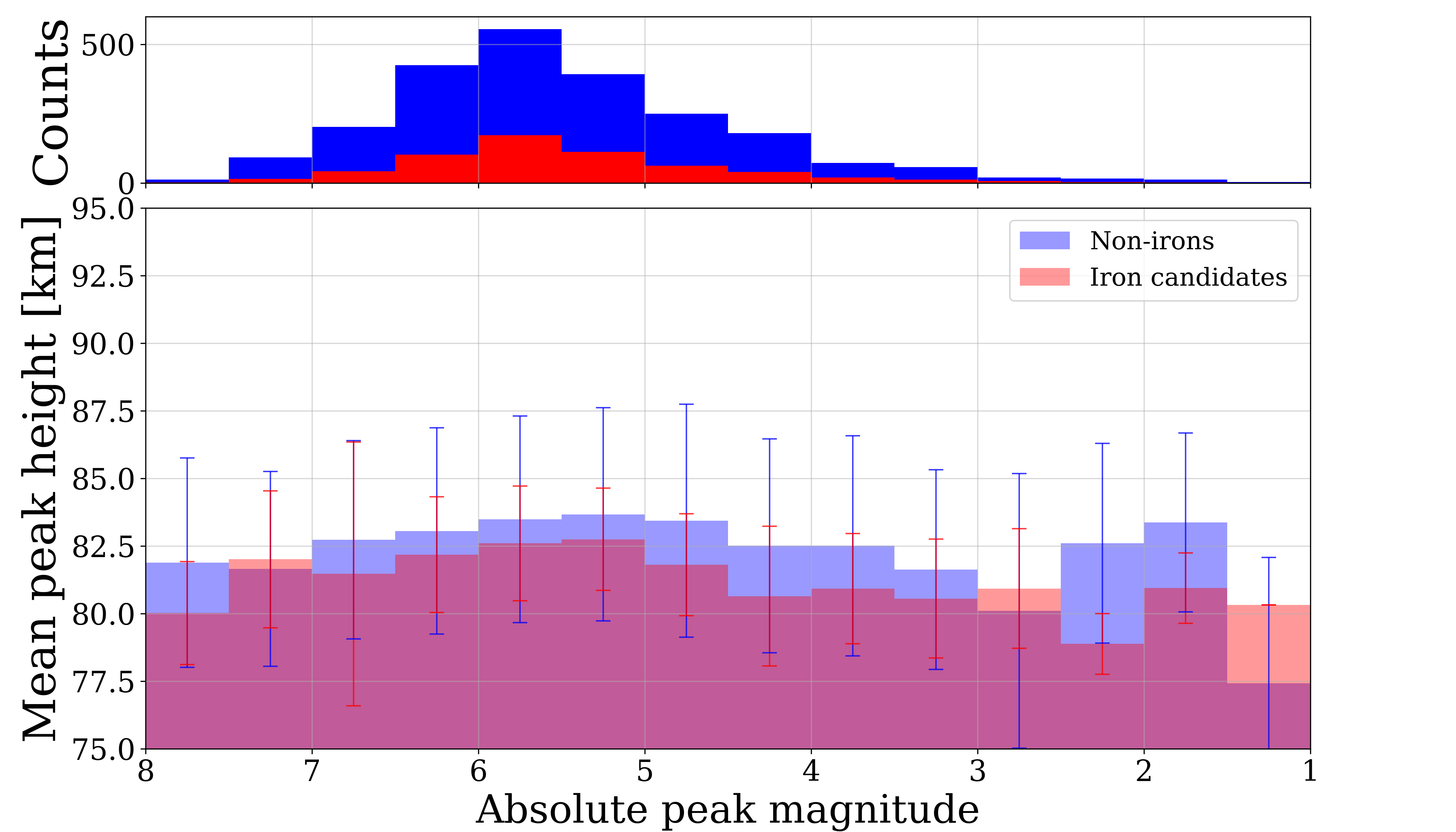}
    \vspace{-0.5cm}
    \caption{Mean peak height versus peak magnitude for iron candidates and non-irons between 10 - 15 km/s, with counts shown above. The difference in peak height between irons and non-irons is consistently between 0.5 - 1.5 km between +7 - +4 magnitudes. The uncertainty bounds represent the standard deviation of the population in a given bin.}
    \label{fig:mean_height_mag}
\end{figure}

To determine if this is a significant bias, in Figure \ref{fig:mean_height_mag}, the mean peak height versus peak magnitude is shown for both iron and non-iron populations. This shows how both populations follow the same peak height trend where they are most numerous. This trend is also almost identical for both median and mean peak heights. The offset in peak height between +7 and +4 magnitudes is about 1 km, which at this height corresponds to only a slightly larger range difference and thus a magnitude difference of $<0.1$. There is little difference in the relative sensitivity of the system to the different populations across these heights.

Since the majority of our iron candidates are at these low speeds (between 10 - 15 km/s), and the weightings of the opposing range and larger mass for deeper penetration biases are small, we ignore these biases when correcting to our final absolute count of iron candidates. The bias corrected fraction of iron candidates can be seen in Table \ref{tab:weighted_counts}.

\begin{figure}
    \centering
    \includegraphics[width=\linewidth]{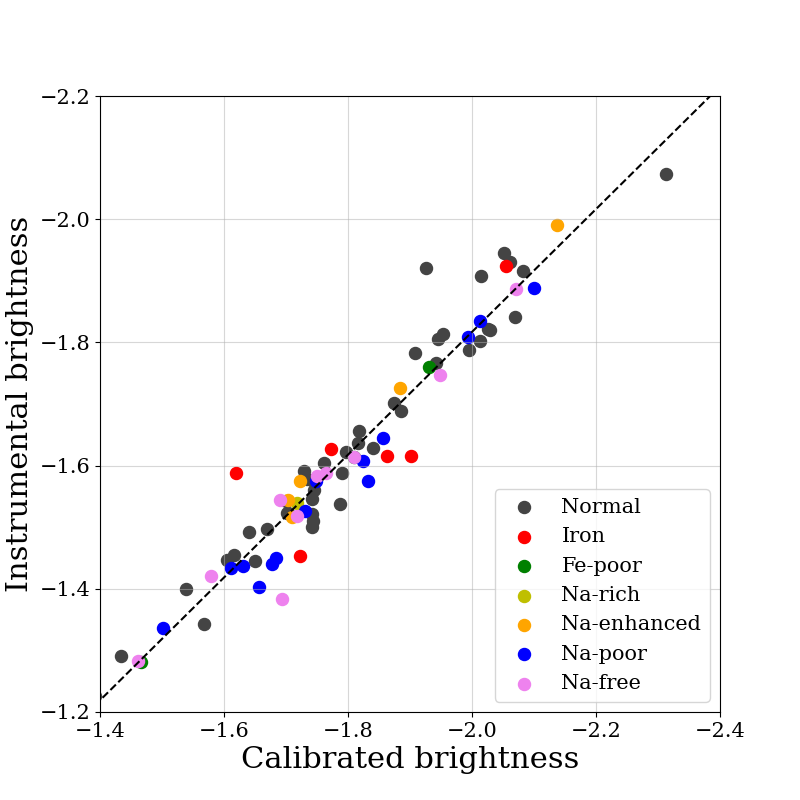}
    \vspace{-0.5cm}
    \caption{The calibrated spectra of \citet{Vojacek2019} are normalized, summed, and compared with the brightness of these spectra convolved with our camera's response function. A photometric comparison line has been fitted with a slope of 1.}
    \label{fig:colour_comparison}
\end{figure}

Another possible source of bias comes from differences in spectra between irons and non-irons. If an iron and non-iron emit the same number of photons but have a completely different spectrum, the observed brightness will depend on our camera's response function\textemdash potentially biasing us towards brighter or fainter meteors. Figure \ref{fig:colour_comparison} shows the calibrated, normalized, and summed brightnesses of meteor spectra from \citet{Vojacek2019} versus these same normalized and summed spectra, but convolved with our camera system's response function, all in arbitrary magnitude units. While there are only a few iron spectra in this dataset, they do fall above and below the centre of this photometric comparison line. This variety suggests that iron meteors are not biased towards brighter or dimmer observed brightness in our EMCCD cameras, and so the weighted counts as calculated above are still robust.

\begin{figure}
    \centering
    \includegraphics[width=\linewidth]{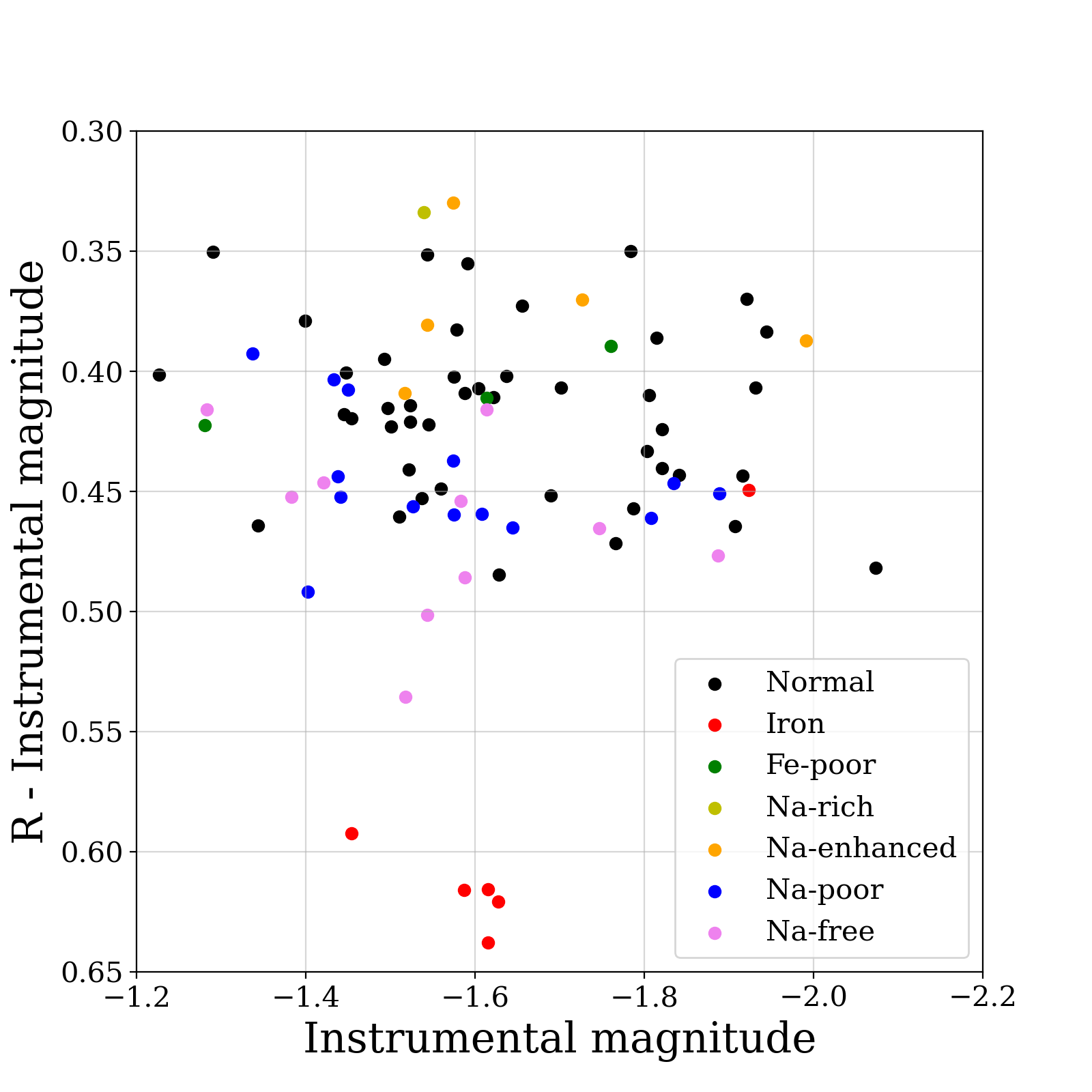}
    \vspace{-0.5cm}
    \caption{The calibrated spectra of \citet{Vojacek2019} normalized, summed, and convolved with our cameras optics on the x-axis, compared with the same convolution plus a standard R colour filter on the y-axis. Iron meteors appear significantly dimmer in the filtered camera as compared to other meteors.}
    \label{fig:colour_distinguish}
\end{figure}

While there is no bias due to the spectra of iron meteors compared to non-iron meteors, we can exploit the differing spectra to better isolate irons via colour \citep{2020JIMO...48..193M}. Using the same data from \citet{Vojacek2019}, a colour comparison can be seen in Figure \ref{fig:colour_distinguish}. Along the x-axis is the instrumental magnitude of these meteors in our camera system, but on the y-axis we also convolve the spectra with a standard Johnsons-Cousins R colour filter and subtract the original instrumental magnitude. The resulting colour clearly distinguishes irons from the general meteor population. Two matching cameras with the same field of view, with one being R-filtered, could theoretically distinguish iron meteors by watching for meteors that are significantly dimmer in the filtered camera. We have begun such observations, but have not yet accumulated enough events to construct a further analysis of the population using this additional discriminator. 

We have opted to leave our interpretations largely to observed brightness values for meteors and not attempted to interpret our results in event-specific mass units and therefore have avoided use of luminous efficiency. The luminous efficiency may be different for irons compared to stony meteoroids at such low speeds, though the luminous efficiency of the latter are usually scaled to that of iron which has better calibration \citep{popova2019}. This could potentially introduce a bias if the iron meteoroid luminous efficiency was much higher than stony meteoroids. However, it is unlikely there could be significant (more than factor of several) difference between the populations as stony meteoroids, presumed to be chemically similar to chondrites, are themselves 20-25\% iron. 

\section{Conclusions}

Our major conclusions include:
\begin{enumerate}

    \item Meteoroids with iron ablation properties are most common at very low entry speeds, comprising $\approx$20\% of the mm-sized meteoroid population with $v<15$ km/s.
    \item Most iron meteoroids in our brightness limited survey ($>95\%$) move in asteroidal orbits; a few percent have Halley-type comet orbits, though some of the latter may be contamination from Na-free meteoroids.
    \item We find that height overlap, range and spectral sensitivity biases are small for our survey and that our raw observed fraction of irons at a given absolute magnitude is within a few percent of the estimated absolute number.
    \item Irons show noticeably lower eccentricities than the equivalent non-iron population between 10-20 km/s. We suggest this may reflect much older collisional ages and significant PR drag effects for irons. 
    \item The observed debiased ratio of iron to non-iron meteoroids has a maximum of 20-25\% for meteors of magnitude +5 - +7 at low speeds ($v<15$ km/s) and shows some evidence of a continuing slow increase right to the brightness limit of our survey.  The ratio decreases somewhat at brighter magnitudes, but remains as high as 15\% at +3.
    \item The origin of the iron population may be metallic chondrules/nodules or some form of impact melt as suggested by \citet{Borovicka2019}. Some may be iron sulfide grains or some sort of stony-iron mix. They appear to be too large to represent typical metal-trolite grains liberated from chondritic parent bodies through collisions \citep{Kuebler1999}.

\end{enumerate}
\label{sec:conclusions}



\section*{Acknowledgements}

This work was supported in part by NASA co-operative agreement 80NSSC21M0073, by the Natural Sciences and Engineering Research Council of Canada Discovery Grants program (Grant no. RGPIN-2018-05659) and the Canada Research Chair Program.

The authors thank Z. Krzeminski and J. Gill for technical support. 

\section*{Data Availability}

The data underlying this article will be shared on reasonable request to the corresponding author.


\bibliographystyle{mnras}
\bibliography{ironrain} 





\bsp	
\label{lastpage}
\end{document}